\documentclass[12pt]{article}
\usepackage{a4wide,feynmp,amsbsy,latexsym,amsfonts,cite}

\newcommand{\no}{\noindent}

\newcommand{\ga}{\gamma}
\newcommand{\de}{\delta}

\newcommand{\cd}{\cdot}

\newcommand{\mod}[1]{\vert {#1}\vert}
\newcommand{\pa}{\partial}
\newcommand{\G}{{\cal G}}

\newcommand{\kb}{{\boldsymbol k}}
\newcommand{\vb}{{\boldsymbol v}}

\newcommand{\intden}{\int\!\frac{d^4k}{(2\pi)^4}}
\newcommand{\ee}{\mathrm{e}}
\newcommand{\ii}{{i}}
\newcommand{\Zmat}[1]{ \sqrt{
\phantom{I}\kern-0.5em\smash{ {\bf
Z}_{#1} } } }
\newcommand{\Zmatb}[1]{\sqrt{
\phantom{I}\kern-0.5em\smash{\overline{{}^{\phantom{{\displaystyle
i}}}\kern-0.4em{\bf Z}_{#1}}}}}

        \newcommand{\zmat}[1]{ \sqrt{
        \phantom{I}\kern-0.5em\smash{ {\mathfrak z}_{#1} } } }
         \newcommand{\zmatb}[1]{\sqrt{
        \phantom{I}\kern-0.5em\smash{\overline{{}^{\phantom{{\displaystyle
        i}}}\kern-0.4em{\mathfrak z}_{#1}}}}}

\newcommand\I{\textsl{I}}

\begin{document}

\begin{titlepage}
\rightline{BNL-HET-99/19} \rightline{PLY-MS-99-24}

\vskip19truemm
\begin{center}{\Large{\textbf{Charges from Dressed Matter:
Physics and Renormalisation}}}\\ [8truemm] \textsc{Emili Bagan\footnote{email:
bagan@quark.phy.bnl.gov; permanent address: IFAE, Univ.~Aut\`onoma de
Barcelona,\\ E-08193 Bellaterra (Barcelona), Spain}\\[5truemm] \textit{Physics
Dept.\\ Brookhaven National Laboratory\\ Upton, NY 11973\\ USA}\\[10truemm]
Martin Lavelle\footnote{email: mlavelle@plymouth.ac.uk}, and David
McMullan}\footnote{email: dmcmullan@plymouth.ac.uk}\\ [5truemm] \textit{School
of Mathematics and Statistics\\ The University of Plymouth\\ Plymouth, PL4
8AA\\ UK} \end{center}

\bigskip\bigskip\bigskip
\begin{quote}
\textbf{Abstract:}
Gauge theories are characterised by long range interactions.
Neglecting these interactions at large times, and identifying the
Lagrangian matter fields with the asymptotic physical fields,
leads to the infra-red problem.
In this paper we study the
perturbative applications of a construction of physical
charges in QED, where the matter fields are
combined with the associated electromagnetic
clouds. This has been formally shown, in a companion paper,
to include these asymptotic interactions.
It is explicitly demonstrated
that the on-shell Green's functions and $S$-matrix elements
describing these charged fields
have, to all orders in the coupling, the pole structure
associated with particle propagation and scattering.
We show in detail that the renormalisation procedure may be
carried out straightforwardly.
It is shown that standard infra-red finite
predictions of QED are not altered
and it is speculated that the good infra-red properties of our
construction may open the way to the calculation of previously
uncalculable properties. Finally
extensions of this approach to QCD are briefly discussed.
\end{quote}

\end{titlepage}

\setlength{\parskip}{1.5ex plus 0.5ex minus 0.5ex}


\section{Introduction}

The description of charged particles is a central task in quantum field
theory. This is obvious for theories such as Quantum Electrodynamics (QED)
and the weak interaction since
particles with such charges, e.g., the electron and the neutrino, are
observed as asymptotic, physical particles in experiments. It is still
a challenge even in Quantum Chromodynamics (QCD), however, since how
the effective quark structure of hadrons emerges from QCD remains an unsolved problem.

In unbroken gauge theories, characterised by massless gauge bosons,
there are long range interactions which, it has long been
known~\cite{Dollard:1964,Chung:1965,kulish:1970},
cannot be neglected even for particles which are a long way from each
other. But if the coupling does not vanish, then the gauge
transformations are non-trivial and, in particular, the matter fields
of QED are not gauge invariant and cannot be identified with physical
particles.  In previous papers we have
shown that by a process of \textit{dressing} the matter fields with a
cloud made out of the vector potentials we can construct fields which
are locally gauge invariant. Many such invariant fields can be
manufactured. The next task is thus twofold: one
must \textit{identify} which of the gauge invariant fields correspond to physically
significant variables and  carry out the \textit{construction} of
these fields.

In a previous paper~\cite{Bagan:1999xx} (henceforth \I)  we argued that a
particular set of gauge invariant fields correspond to
physical charges in QED. In particular we showed that the slow
fall-off of the  interaction in QED, which is itself a
consequence of the masslessness of the gauge boson, means that this
interaction cannot be
neglected even for charges which are highly separated from each
other and that it is not justified to set the QED coupling to zero at
large times in scattering processes. These unjustified
assumptions generate the infra-red (IR) problem.
We demonstrated that the creation
and annihilation operators do create physical (gauge invariant)
particles (which even have the expected electromagnetic fields!), but
that these operators do \emph{not} correspond to the asymptotic limits
of the Lagrangian matter field. We were able to obtain the gauge
invariant fields which do, at the correct points on the mass shell,
yield physical particles obeying a free asymptotic dynamics. We
further presented an equation
with which we could generate these dressed
fields corresponding to physical charges. For the purposes of this
paper, we may summarise the
results of \I\ in one equation: we claim\footnote{Note that in this paper
we neglect phase effects which play no physical  role in QED.} that the locally gauge
invariant field
\begin{equation}\label{theone}
 \psi_v(x):= h^{-1}(x)\psi(x)=\ee^{-ie\chi(x)}\psi(x)\,,
\end{equation}
where
\begin{equation}\label{4min}
  \chi(x)=\frac{\G\cd A}{\G\cd\pa}\,,
\end{equation}
with $\G^\mu=(\eta+v)^\mu(\eta-v)\cd\pa-\pa^\mu$,
\emph{corresponds to a charged particle moving with four-velocity},
$u=\gamma(\eta+v)$ where $\eta$ is the time-like unit vector,
$v=(0,\vb)$ is the velocity and $\gamma$ is the usual Lorentz
 factor, $\gamma=1/\sqrt{1-\vb^2}$. We should reiterate
that we may \emph{only} interpret these fields as describing physical
particles at the on-shell point $p=m\gamma(\eta+v)$.

This paper is dedicated to testing in detail
the results of \I\ in the context of
perturbative calculations in QED. In particular we
will study the infra-red  and ultra-violet (UV)
properties of the dressed fields in (\ref{theone}).
It will be shown that the IR divergences associated with going
on-shell are indeed
removed already at the level of S-matrix elements at all
orders in perturbation theory. (Some similar results for scalar QED
have been
announced elsewhere \cite{Bagan:1998kg,Horan:1998im}; here we consider the
fermionic theory and give details of our calculations and techniques.)
Then the UV structure of these fields will be investigated and
it will be shown that the dressed fields have
a good UV behaviour.

We note that there is, however, in QCD a general non-perturbative obstruction
to this construction of effective quarks and this sets a fundamental
limit on the validity of quark models \cite{Lavelle:1997ty}.

As well as identifying the above mentioned limit on the construction
of observable quarks and constructing  solutions for
dressed electrons, quarks and gluons (the last two in perturbation
theory) which have  a specific physical identification,
the notable successes of this programme include the demonstration,
in QED, that the dressed Green's functions for
charge propagation possess a good pole
structure~\cite{Bagan:1997dh,Bagan:1997su},
that $S$-matrix elements for interactions with
dressed asymptotic fields are free of on-shell IR divergences
\cite{Bagan:1998kg} and that
the paradigm anti-screening behaviour of QCD can be understood in
terms of the interaction between two separately gauge-invariant
coloured quarks \cite{Lavelle:1998dv}.

The structure of this paper is as follows: after outlining the
Feynman rules in Sect.\ 2,
(in particular for interactions with the dressing cloud)
we study the IR behaviour of the Green's functions
of dressed fields in Sect.\ 3 to all orders in perturbation theory.
Then, in Sect.\ 4, the UV sector comes
under attack: it is shown that the renormalisation of the charged
fields, understood here as composite operators, is straightforward and
that they do not mix with other operators under renormalisation. Other
important properties, such as Ward identities and a demonstration that using physical
variables yields the correct result for $g-2$, are also proven here.
Finally in Sect.\ 5 our results are summarised and conclusions are
drawn. Various calculational details are given in appendices.

\begin{fmffile}{af}

\fmfcmd{%
vardef port (expr t, p) =
 (direction t of p rotated 90)
  / abs (direction t of p)
 enddef;}

\fmfcmd{%
  vardef portpath (expr a, b, p) =
   save l; numeric l; l = length p;
    for t=0 step 0.1 until l+0.05:
    if t>0: .. fi point t of p
shifted  ((a+b*sind(180t/l))*port(t,p))  endfor  if cycle  p:  ..
cycle fi enddef;}

\fmfcmd{%
  style_def brown_muck expr p =
   shadedraw(portpath(thick/2,2thick,p)
    ..reverse(portpath(-thick/2,-2thick,p))
    ..cycle)
  enddef;}

\section{Feynman rules}\label{s1}

In order to perform perturbative calculations, we first need to
know
the Feynman rules.
For the standard propagators and vertices we
stick to the conventions used in Bjorken and Drell except for the
sign of the electric charge, $e$, which we choose to be the
opposite. In this section we briefly outline the derivation of
the new set of Feynman rules for the insertions of the
dressed electron, which we then summarise
in Fig.~\ref{f3}.
\begin{figure}[tbp]
$$
\begin{array}{lclcl}
\parbox{20mm}{
\begin{fmfgraph}(20,5)
\fmfleft{l}
\fmfright{r}
\fmf{fermion}{l,r}
\fmfv{d.sh=c,d.filled=empty,d.si=2mm}{r}
\end{fmfgraph}}
&=&
\parbox{20mm}{
\begin{fmfgraph}(20,5)
\fmfleft{l}
\fmfright{r}
\fmf{fermion}{r,l}
\fmfv{d.sh=c,d.filled=empty,d.si=2mm}{r}
\end{fmfgraph}}
&=&1\\[0.5cm]
\parbox{20mm}{
\begin{fmfgraph*}(20,20)
\fmfstraight
\fmfbottomn{l}{2}
\fmftopn{r}{4}
\fmf{fermion}{l1,l2}
\fmf{photon,label=$k_{1}\searrow$,label.side=left,l.d=.5mm}{l2,v1}
\fmf{phantom,tension=3}{v1,r1}
\fmf{photon,label=$\downarrow \kern1mm k_{n}$,l.side=left}{r4,l2}
\fmfv{d.sh=c,d.filled=empty,d.si=2mm}{l2}
\fmf{photon}{l2,v2}
\fmf{phantom,tension=8}{v2,r2}
\fmf{dots,left=.2}{v2,r4}
\end{fmfgraph*}}
\kern 1.5cm
&=&
\parbox{20mm}{
\begin{fmfgraph*}(20,20)
\fmfstraight
\fmfbottomn{l}{2}
\fmftopn{r}{4}
\fmf{fermion}{l2,l1}
\fmf{photon,label=$k_{1}\nwarrow$,label.side=left,l.d=.5mm}{l2,v1}
\fmf{phantom,tension=3}{v1,r1}
\fmf{photon,label=$\uparrow \kern1mm k_{n}$,l.side=left}{r4,l2}
\fmfv{d.sh=c,d.filled=empty,d.si=2mm}{l2}
\fmf{photon}{l2,v2}
\fmf{phantom,tension=8}{v2,r2}
\fmf{dots,left=.2}{v2,r4}
\end{fmfgraph*}}
\kern 1.5cm
&=&\displaystyle
{eV_{1}^{\mu_{1}}\over V_{1}\cdot k_{1}}{eV_{2}^{\mu_{2}}\over V_{2}\cdot
k_{2}}\cdots
{eV_{n}^{\mu_{n}}\over V_{n}\cdot k_{n}}
\end{array}
$$
        \caption{Feynman rules for the insertions of the dressing. We have
        used the notation $V_{r}^{\mu_{r}}=(\eta+v)^{\mu_{r}}(\eta-v)\cdot
        k_{r}-k_{r}^{\mu_{r}}$, where $\eta^\mu=(1,{\bf 0})$ and
$v^\mu=(0,\vb)$.
        Note that $v^2=-\vb^2$.}
        \label{f3}
\end{figure}
We will sometimes refer to these extra vertices, typical of the physical Green's
functions, as \emph{dressing vertices}.
We recall from the introduction that our dressed
matter  $\psi_v={\rm e}^{-{\rm i} e\chi}\psi$
has an explicit dependence on the coupling constant.
We therefore expand the exponential and insert each term, $(-{\rm i}e \chi)^n/n!$,
of the expansion in
the appropriate Green's function (one with a $\bar\psi$ field and $n$
photon fields).
The $n!$ is seen to cancel upon contracting the photon fields.
According to the standard
procedure, we Fourier transform the result and amputate the external
propagators.
The final results in Fig.~\ref{f3} can be read off from the expressions
for  ${\chi}$ and ${\cal G}_\mu$ in~(\ref{4min}). Note that
the vectors $V_l$, defined
in the caption of Fig.~\ref{f3}, are essentially the Fourier transforms
of ${\cal G}$
with respect to the variables~$k_l$. One should also be aware that
diagrams such as that in Eq.~\ref{eb1b}
of Appendix~\ref{fact_app} have a statistical factor $1/n!$, where $n$
is the number of
internal photon lines whose ends are \emph{both} attached to dressing
vertices ($n=2$ in that equation). These lines
will be called {\em rainbow lines} later on.


\newcommand{\slsh}{\kern-0.5em/}
\section{Dressed Matter in the Infra-Red}\label{thr}
\setlength{\unitlength}{1mm}

\no In this section we wish to study the effects of
dressings upon the soft or infra-red structure of gauge theories.
IR singularities in gauge theories  generally  prevent us from
 following the usual LSZ route to
the $S$-matrix, however, in \I\ we argued that it should be possible
to pursue the LSZ path using dressed fields. We further recall
that in previous papers we have seen that the on-shell dressed
matter propagator is free of IR divergences
if the dressing is appropriate to the point on the mass shell where we
renormalise (e.g., if we renormalise at $p=(m,0,0,0)$, then the dressing
must be that appropriate to a static charge). This has been seen for a
general relativistic charge for both the fermionic \cite{Bagan:1998kg}
and scalar theories \cite{Bagan:1997dh}. We have
elsewhere \cite{Bagan:1998kg}
sketched a proof that, in
scalar QED, using the appropriate dressings on the various
charged external legs, the  IR divergences associated with going
on shell cancel for both scattering and pair production in dressed
Green's functions and the $S$-matrix. Here we will demonstrate that
this holds for the
fermionic theory and provide the full details which were not given in
Ref.~\cite{Bagan:1998kg}. In particular, we will prove
the property of factorisation of
rainbow lines which was
stated without proof in our earlier paper.

We start this section  with a study of the one-loop Green's functions of
our dressed charges. We will here study the contribution of soft momenta
in loop integrals, since only they can generate IR divergences, and,
to this purpose, we
introduce a refined method with which we can rapidly show the cancellation
of on-shell IR divergences. All terms which cannot generate IR problems
will be
systematically discarded. (This is of course a gauge invariant procedure
since our physical charged fields are themselves by construction gauge
invariant.) We shall
apply this technique to  the propagator and also
to scattering off a source. It should become clear that this cancellation
only holds for charged fields taken at the appropriate
points on the mass shell.
We then move on to an all orders study. Here we discuss the form
of the propagator which tells us about the structure of the
renormalisation constants. It is shown that the IR divergences in
the $S$-matrix elements which describe charged matter scattering
off a source\footnote{This
and the one-loop arguments can be easily generalised to
other scattering processes.} exponentiate and, from our one-loop
result, this implies that they are IR finite at all orders in
perturbation theory. Since the on-shell renormalisation
constants for our dressed fields are IR finite, these results for
$S$-matrix elements are then straightforwardly carried over to
the dressed Green's functions at the right points on the mass shells.
These arguments make extensive use of the factorisation property shown
in Appendix~\ref{fact_app}.

\subsection{One-Loop Green's Functions}
\label{IR1}

\no To calculate $S$-matrix elements for some general process, we need
to calculate on-shell Green's functions and amputate a leg for each
external particle. As is well known, the IR problem reveals itself as
divergences in the residues of these poles.
We want here to calculate {\em dressed} Green's functions and show
that the dressings generate new terms which exactly cancel the usual
IR divergences, i.e., we demonstrate that $S$-matrix elements
constructed in terms of dressed charges are IR finite.

Although below we will see that we do not even need to perform the
integrals explicitly to see the cancellation of on-shell IR
divergences in dressed Green's functions and the corresponding
$S$-matrix elements, we should briefly discuss how such divergences
may be handled. It is clear that we want to respect gauge invariance
and dimensional regularisation is therefore a natural regularisation
scheme for us. Here \cite{Bagan:1997su} the IR
divergences show up as divergences in
Feynman parameter integrals which yield poles in $1/(4-D)$. Another
method which may be used is that described in Chapter 13
of \cite{weinberg:1995}, here we look for poles in $1/k^2$, use the
residue theorem to integrate over $k_{0}$ and so obtain
IR singularities as divergences at the lower limit ($\lambda$) of
the integral over $\mod {\kb}$. A variation on this last
scheme keeps the charged matter slightly off shell,
$\Delta=p^2-m^2\neq0$, which yields logarithms in $\Delta$. This last
approach brings out the primarily on-shell nature of the IR problem. The
results of all these methods may, however, be straightforwardly translated
into each other with the help of the following \lq dictionary\rq\
\begin{equation}
-\log\frac{\Lambda^2}{p^2-m^2}\leftrightarrow
\log\frac{\Lambda}{\lambda}\leftrightarrow
\int^1_{0} du u^{D-5}=\frac1{D-4}
\,,
\label{e-20.7a}
\end{equation}
where the form of $\Lambda$ is irrelevant for the singularity.
Since, as far as this section is concerned, we can perform all of the
necessary manipulations at the level of integrands, we do not need to
explicitly perform the integrations -- as such we simply write our
inetegrals over four-dimensional momenta. All the above schemes,
though, may be directly applied.

\no \textbf{a) Propagating Charges}

\smallskip

\begin{figure}[tbp]
\begin{eqnarray*}
&
         \parbox{20mm}{
\begin{fmfgraph*}(20,20)
\fmfleft{l}
\fmfright{r}
\fmf{fermion,label=(a),l.d=0.5cm}{l,r}
\fmfv{d.sh=c,d.filled=empty,d.si=2mm}{r}
\fmfv{d.sh=c,d.filled=empty,d.si=2mm}{l}
\end{fmfgraph*}
}
\quad
+ &\\
&
        \parbox{20mm}{
\begin{fmfgraph*}(20,20)
\fmfleft{l}
\fmfright{r}
\fmf{fermion}{l,v,r}
\fmffreeze
\fmf{photon,left=1}{v,r}
\fmfv{d.sh=c,d.filled=empty,d.si=2mm}{r}
\fmfv{d.sh=c,d.filled=empty,d.si=2mm}{l}
\fmfv{d.sh=c,d.filled=empty,d.si=0mm,label=(b),l.d=0.5cm,l.a=-90}{v}
\end{fmfgraph*}
}
\quad
+\quad
        \parbox{20mm}{
\begin{fmfgraph*}(20,20)
\fmfleft{l}
\fmfright{r}
\fmf{fermion}{l,v,r}
\fmffreeze
\fmf{photon,left=1}{l,v}
\fmfv{d.sh=c,d.filled=empty,d.si=2mm}{r}
\fmfv{d.sh=c,d.filled=empty,d.si=2mm}{l}
\fmfv{d.sh=c,d.filled=empty,d.si=0mm,label=(c),l.d=0.5cm,l.a=-90}{v}
\end{fmfgraph*}
}
\quad
+\quad
        \parbox{20mm}{
\begin{fmfgraph*}(20,20)
\fmfleft{l}
\fmfright{r}
\fmf{fermion,label=(d),l.d=0.5cm}{l,r}
\fmffreeze
\fmf{photon,left=0.5}{l,r}
\fmfv{d.sh=c,d.filled=empty,d.si=2mm}{r}
\fmfv{d.sh=c,d.filled=empty,d.si=2mm}{l}
\end{fmfgraph*}
}
\quad
+\quad
        \parbox{20mm}{
\begin{fmfgraph*}(20,20)
\fmfleft{l}
\fmfright{r}
\fmf{fermion}{l,v1}
\fmf{fermion,tension=1,label=(e),l.d=0.5cm}{v1,v2}
\fmf{fermion}{v2,r}
\fmffreeze
\fmf{photon,left=1}{v1,v2}
\fmfv{d.sh=c,d.filled=empty,d.si=2mm}{r}
\fmfv{d.sh=c,d.filled=empty,d.si=2mm}{l}
\end{fmfgraph*}
}
&
\end{eqnarray*}
        \caption{One-loop diagrams contributing to the
        propagator ${i}S_{v}(p)$. Note that we display a circle
        vertex at the end of the
fermion lines to show that this is a dressed fermion: at order $g^0$
this vertex reduces to unity, but we wish to distinguish it from the
usual fermion. This distinction will be more important in the next
section.}
        \label{f4}
\end{figure}
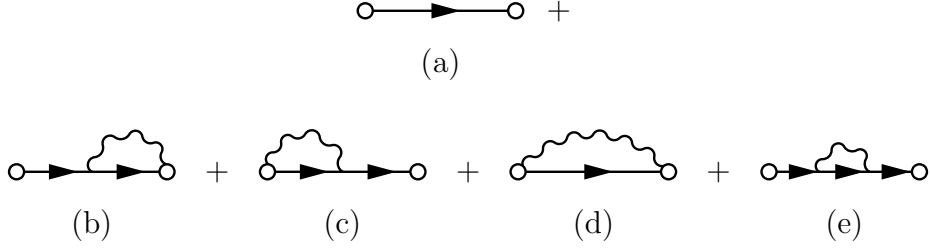

\no Let us initially consider the simplest Green's function --- that
describing a single propagating charge. The relevant diagrams
are shown in Fig.~\ref{f4}. It is well known that
the usual on-shell propagator
(the sum of diagrams \ref{f4}a and \ref{f4}e) displays a soft divergence.
The propagator of the dressed charge, defined as
\begin{equation}
i S_v(p)\equiv \langle \psi_v(p) \bar\psi_v(p)\rangle
\equiv\int{{d}^4 x\over(2\pi)^4}\;\ee^{ i p\cdot x}
\langle 0|{\rm T} \psi_v(x) \bar\psi_v(0)|0\rangle
,
\end{equation}
receives at order $e^2$
further corrections from the Feynman rules corresponding to the expansion
of the dressings. These terms are represented diagramatically at one loop
by the diagrams \ref{f4}b--d.

The contribution of diagram \ref{f4}b to the propagator has,
from our Feynman rules, the form
\begin{equation}
iS_{v}^{\ref{f4}\mathrm{b}}(p)=-e^2\intden
\frac{V^{\mu}}{V\cdot k}\frac{p\slsh - k
\slsh+m}{(p-k)^2-m^2}\gamma^{\nu} D_{{\mu\nu}}(k)\frac{1}{p\slsh-m}
\,,
\end{equation}
where, to help us bring out the gauge invariance
of our final result below, we have
left the form of the vector boson propagator, $D_{\mu\nu}$,
completely general (in Feynman gauge, we would have $D_{\mu\nu}=-
g_{\mu\nu}/k^2$). The definition of $V$ and its dependence on $v$
should be obvious, cf.\ Sect.~2.
We now drop the term proportional to $k$ in the numerator, as it will
be IR finite. We may rewrite $(p\slsh+m)\gamma^{\nu}$ as $2p^{\nu}-
\gamma^{\nu}(p\slsh-m)$, and
to calculate the on-shell
residue of the pole we only have to retain $2p^{\nu}$. This
has then the form
\begin{equation}
iS_{v}^{\ref{f4}\mathrm{b}}(p)=\frac{e^2}{p\slsh-m}
\intden \frac{V^{\mu}p^{\nu}}{V\cdot k p\cdot k}D_{{\mu\nu}}(k)
\,,
\end{equation}
which, we see by power counting, has an IR divergence. The
contribution of diagram \ref{f4}c is easily seen to be identical to this.

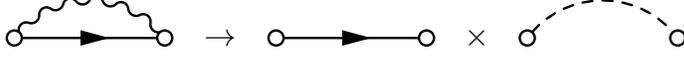
\begin{figure}[tbp]
$$
        \parbox{20mm}{
\begin{fmfgraph*}(20,20)
\fmfleft{l}
\fmfright{r}
\fmf{fermion,label=$\phantom{2}$}{l,r}
\fmffreeze
\fmf{photon,left=0.5}{l,r}
\fmfv{d.sh=c,d.filled=empty,d.si=2mm}{r}
\fmfv{d.sh=c,d.filled=empty,d.si=2mm}{l}
\end{fmfgraph*}
}
\quad
\rightarrow
\quad
         \parbox{20mm}{
\begin{fmfgraph*}(20,20)
\fmfleft{l}
\fmfright{r}
\fmf{fermion,label=$\phantom{2}$}{l,r}
\fmfv{d.sh=c,d.filled=empty,d.si=2mm}{r}
\fmfv{d.sh=c,d.filled=empty,d.si=2mm}{l}
\end{fmfgraph*}
}
\quad
\times
\quad
        \parbox{20mm}{
\begin{fmfgraph*}(20,20)
\fmfleft{l}
\fmfright{r}
\fmf{phantom}{l,r}
\fmffreeze
\fmf{dashes,label=$\phantom{2}$,left=0.5}{l,r}
\fmfv{d.sh=c,d.filled=empty,d.si=2mm}{r}
\fmfv{d.sh=c,d.filled=empty,d.si=2mm}{l}
\end{fmfgraph*}
}
$$
        \caption{Diagrammatic representation of the factorisation of
        the one-loop propagator rainbow diagram.}
        \label{appa2}
\end{figure}

As far as diagram \ref{f4}d is concerned, the Feynman rules yield
\begin{equation}
iS_{v}^{\ref{f4}\mathrm{d}}(p)=e^2\intden \frac{1}{p\slsh -
k\slsh -m}
\frac{V^{\mu}V^{\nu}}{V\cdot k V\cdot k}D_{{\mu\nu}}(k)
\,.
\end{equation}
In such \textit{rainbow} diagrams, where  one or more
vector boson lines extend from one
dressing vertex to another,  we can, as is shown in
Appendix~\ref{fact_app},  so far as the infra-red or
ultra-violet structure is concerned, always factor off the loop
integrals associated with such lines to obtain a
product of the underlying Feynman diagram without the rainbow
lines (here the
free fermion propagator) times the loop integrals; this is
represented diagrammatically in Fig.~\ref{appa2}.
To do this
we make use of Eq.~\ref{basic} from Appendix~\ref{fact_app} to
rewrite this as
\begin{equation}
iS_{v}^{\ref{f4}\mathrm{d}}(p)=\frac{-e^2}{p\slsh-m}\intden
\frac{V^{\mu}V^{\nu}}{V\cdot k \, V\cdot k}D_{{\mu\nu}}(k)
\,.
\label{e-19.7b}
\end{equation}

Finally we need to find the contribution of the usual covariant
diagram, \ref{f4}e. From the Feynman rules we obtain
\begin{equation}
iS_{v}^{\ref{f4}\mathrm{e}}(p)=\frac{e^2}{p\slsh-m}\intden \gamma^{\mu}
\frac1{p\slsh+k\slsh-m}\gamma^{\nu}
\frac{1}{p\slsh-m } D_{{\mu\nu}}(k)
\,,
\end{equation}
which appears to have a double pole and to be IR finite. This double pole
will be killed off by mass renormalisation.
However, we
can extract a power of $(p^2-m^2)$ and in the process also
obtain a single pole with an IR
divergence.
Proceeding now as with diagram \ref{f4}b, we may rewrite this as
\begin{equation}
iS_{v}^{ \ref{f4}\mathrm{e}}(p)=\frac{e^2}{p\slsh-m}\intden \gamma^{\mu}
\frac{2p^\nu}{[(p-k)^2-m^2]} \frac{1}{p\slsh-m}D_{{\mu\nu}}(k)
\,,
\end{equation}
plus terms with a single pole which are IR finite and hence
not considered  here. If we expand
the integral
in $(p^2-m^2)$ we obtain
\begin{equation}
iS_{v}^{ \ref{f4}\mathrm{e}}(p)=-\frac{e^2}{p\slsh-m}\intden \gamma^{\mu}
\frac{2p^\nu (p^2-m^2)}{(2p\cdot k)^{2}}
\frac{1}{p\slsh-m}D_{{\mu\nu}}(k)
\,.
\end{equation}
where we have also dropped the non-pole like,
IR finite terms (i.e., everything except the simple pole).
We now write $(p^2-m^2)=
(p\slsh+m)(p\slsh-m)$: the latter factor here removes the
double pole and the first factor can be taken through the remaining
gamma matrix to yield effectively $2p^{\mu}$ (plus
terms which do not contain a pole in the propagator). In this way we
obtain
\begin{equation}
iS_{v}^{\ref{f4}\mathrm{e}}(p)=-\frac{e^2}{p\slsh-m}\intden
\frac{p^\mu p^{\nu}}{(p\cdot k)^{2}} D_{{\mu\nu}}(k)
\,.
\end{equation}
We now combine the results for the various diagrams  to obtain the following form
for the IR divergent terms in the residue of the simple pole in the
dressed matter propagator:
\begin{equation}\label{cgf}
iS^{\mathrm{IR}}_{v}(p)=-\frac{e^2}{p\slsh-m}\intden
\left[
\frac{p^\mu}{p\cdot k} -\frac{V^\mu}{V\cdot k}\right]
D_{{\mu\nu}}(k)
\left[
\frac{p^\nu}{p\cdot k} -\frac{V^\nu}{V\cdot k}\right]
\,.
\end{equation}
This form confirms the gauge invariance of our construction: any photon
propagator other than that of Feynman gauge will involve either a
$k_\mu$ or a $k_\nu$ factor, but these extra terms  will vanish on multiplying
into the square brackets in (\ref{cgf}).

Before showing that this divergence cancels, we should like to
introduce our renormalisation conventions.
From previous work~\cite{Bagan:1997su} we know
that the wave
function renormalisation constant for dressed fermions is in
general a matrix. For the scalar theory this is naturally not the
case, and we stress that for both the scalar~\cite{Bagan:1997dh} and
fermionic~\cite{Bagan:1997su}
theories we were able to multiplicatively
renormalise the
propagator of dressed matter moving at an arbitrary relativistic
velocity. We will return to this in
Sect.~\ref{sec4} and generalise these results. We define a wave-function
matrix renormalisation constant, $\Zmat v$, for
charged matter moving with four-velocity $\gamma(\eta+v)$:
\begin{equation}\label{Zmat}
\psi^{\mathrm B}_v=\Zmat v\psi^{\mathrm R}_v\quad
\Rightarrow \quad
S^{\mathrm B}_v = \Zmat v\, S^{\mathrm R}_v \Zmatb v
\,,
\end{equation}
where $\Zmatb v$ is the adjoint (we stress that the matrix is
defined as $\Zmat v$ for
familiarity and is \emph{not} to be understood as the square root
of some matrix) and $\psi^{\mathrm B}_v$ and $\psi^{\mathrm R}_v$
are the bare and renormalised physical charged fields.

From (\ref{Zmat}) and (\ref{cgf}) it follows that the IR
contribution to the on-shell wave function renormalisation constant
is given by
\begin{equation}
{\Zmat v}^{\mathrm{IR}}=
i\frac{e^2}2\intden
\left[
\frac{p^\mu}{p\cdot k} -\frac{V^\mu}{V\cdot k}\right]
D_{{\mu\nu}}(k)
\left[
\frac{p^\nu}{p\cdot k} -\frac{V^\nu}{V\cdot k}\right]
\,,
\end{equation}
evaluated at that $p$ corresponding to the correct point on the
mass-shell, i.e., $p=m\gamma(\eta+v)$.

That this sum of the different IR divergences
actually vanishes completely may be seen either by
a brute force calculation or, alternatively, by realising that the term
\begin{equation}
\frac{V^\mu}{V\cdot k}=
\frac{(\eta+v)^\mu(\eta-v)\cdot k -k^\mu}{(k\cdot \eta)^2-{k}^2
-({k}\cdot {v})^2}
\,,
\end{equation}
can, for the soft divergences,
i.e., the region where $k^2\approx0$, be rewritten as
\begin{equation}
\frac{V^\mu}{V\cdot k}=
\frac{(\eta+v)^\mu(\eta-v)\cdot k -k^\mu}{(k\cdot \eta)^2
-({k}\cdot{v})^2}
\,,
\end{equation}
and we may now also drop the
$k^\mu$ term in
the numerator  here  (cf,  the
argument  for  the  gauge  invariance  of  Eq.~\ref{cgf}).  So
effectively we obtain
\begin{equation}\label{mayodiez}
\frac{V^\mu}{V\cdot k}=
\frac{(\eta+v)^\mu(\eta-v)\cdot k}{(k\cdot \eta)^2
-({k}\cdot{v})^2}
=\frac{(\eta+v)^\mu}{(\eta+v)\cdot k}
=\frac{p^\mu}{p\cdot k}
\,,
\end{equation}
where this last step in the effective replacement, $p^\mu/p\cdot
k \leftrightarrow V^\mu/V\cdot k$, \emph{only} holds if
we renormalise at the appropriate point on the mass shell,
i.e., $p=m\gamma(\eta+v)$. In this fashion we
may read off the cancellation
of the IR divergences in the on-shell propagator at one loop. We
stress that if we renormalise at a different point on the mass
shell, then the IR divergences are \emph{not} cancelled. (This argument
is closely related to the one used in Sect.'s~3 and 4 of \I\ to show that the
distortion factor is cancelled for our dressed matter.)

We further note that (\ref{cgf}) is essentially identical to
the result \cite{Bagan:1997dh,Horan:1998im} obtained for the scalar
theory (once $1/(p\slsh-m)$ is
replaced by $1/(p^2-m^2)$. This brings out
the well known spin-independence of the IR divergences in theories with
massive charges.

Having thus shown the cancellation of the various IR
divergences which occur in the residue of
the propagator evaluated in
wave-function, on-shell renormalisation, let us
move on to a charge being scattered off a source.

\no \textbf{b) Scattering Charges}

\smallskip
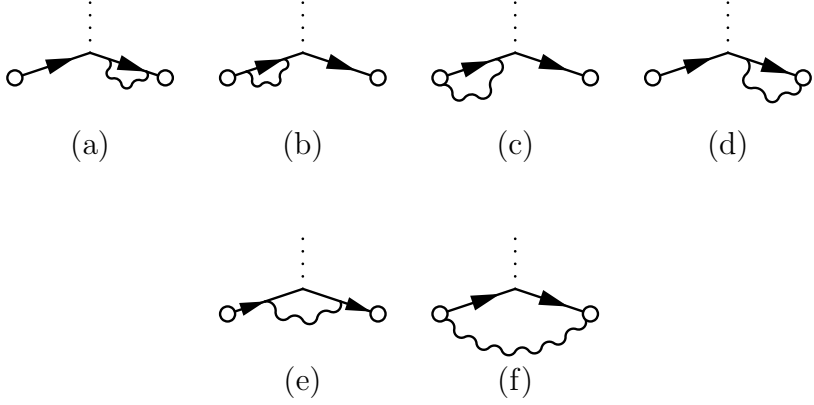
\begin{figure}[tbp]
\begin{eqnarray*}
&\displaystyle
 \parbox[b]{20mm}{
\begin{fmfgraph*}(20,10)
\fmftopn{t}{3}
\fmfbottomn{b}{3}
\fmf{phantom}{t2,vt1}
\fmf{phantom,tension=1/2}{vt1,vt2}
\fmf{phantom}{vt2,v}
\fmf{phantom}{b1,vl1}
\fmf{phantom,tension=1/2}{vl1,vl2}
\fmf{phantom}{vl2,v}
\fmf{phantom}{v,vr1}
\fmf{phantom,tension=1/2}{vr1,vr2}
\fmf{phantom}{vr2,b3}
\fmffreeze
\fmf{fermion}{vr1,vr2}
\fmf{fermion}{b1,v}
\fmf{plain}{v,vr1}
\fmf{plain}{vr2,b3}
\fmfv{d.sh=c,d.filled=empty,d.si=2mm}{b3}
\fmfv{d.sh=c,d.filled=empty,d.si=2mm}{b1}
\fmf{dots}{t2,v}
\fmf{photon,right=1}{vr1,vr2}
\fmf{phantom,label=(a),l.dis=0.7cm}{b1,b3}
\end{fmfgraph*}}
\qquad
        \parbox[b]{20mm}{
\begin{fmfgraph*}(20,10)
\fmftopn{t}{3}
\fmfbottomn{b}{3}
\fmf{phantom}{t2,vt1}
\fmf{phantom,tension=1/2}{vt1,vt2}
\fmf{phantom}{vt2,v}
\fmf{phantom}{b1,vl1}
\fmf{phantom,tension=1/2}{vl1,vl2}
\fmf{phantom}{vl2,v}
\fmf{phantom}{v,vr1}
\fmf{phantom,tension=1/2}{vr1,vr2}
\fmf{phantom}{vr2,b3}
\fmffreeze
\fmf{fermion}{vl1,vl2}
\fmf{fermion}{v,b3}
\fmf{plain}{b1,vl1}
\fmf{plain}{vl2,v}
\fmfv{d.sh=c,d.filled=empty,d.si=2mm}{b3}
\fmfv{d.sh=c,d.filled=empty,d.si=2mm}{b1}
\fmf{dots}{t2,v}
\fmf{photon,right=1}{vl1,vl2}
\fmf{phantom,label=(b),l.dis=0.7cm}{b1,b3}
\end{fmfgraph*}}
\qquad
       \parbox[b]{20mm}{
\begin{fmfgraph*}(20,10)
\fmftopn{t}{3}
\fmfbottomn{b}{3}
\fmf{phantom}{t2,vt1}
\fmf{phantom,tension=1/2}{vt1,vt2}
\fmf{phantom}{vt2,v}
\fmf{phantom}{b1,vl1}
\fmf{phantom,tension=1/2}{vl1,vl2}
\fmf{phantom}{vl2,v}
\fmf{phantom}{v,vr1}
\fmf{phantom,tension=1/2}{vr1,vr2}
\fmf{phantom}{vr2,b3}
\fmffreeze
\fmf{fermion}{v,b3}
\fmf{fermion}{b1,vl2}
\fmf{plain}{vl2,v}
\fmfv{d.sh=c,d.filled=empty,d.si=2mm}{b3}
\fmfv{d.sh=c,d.filled=empty,d.si=2mm}{b1}
\fmf{dots}{t2,v}
\fmf{photon,right=1}{b1,vl2}
\fmf{phantom,label=(c),l.dis=0.7cm}{b1,b3}
\end{fmfgraph*}}
\qquad
        \parbox[b]{20mm}{
\begin{fmfgraph*}(20,10)
\fmftopn{t}{3}
\fmfbottomn{b}{3}
\fmf{phantom}{t2,vt1}
\fmf{phantom,tension=1/2}{vt1,vt2}
\fmf{phantom}{vt2,v}
\fmf{phantom}{b1,vl1}
\fmf{phantom,tension=1/2}{vl1,vl2}
\fmf{phantom}{vl2,v}
\fmf{phantom}{v,vr1}
\fmf{phantom,tension=1/2}{vr1,vr2}
\fmf{phantom}{vr2,b3}
\fmffreeze
\fmf{fermion}{vr1,b3}
\fmf{fermion}{b1,v}
\fmf{plain}{v,vr1}
\fmfv{d.sh=c,d.filled=empty,d.si=2mm}{b3}
\fmfv{d.sh=c,d.filled=empty,d.si=2mm}{b1}
\fmf{dots}{t2,v}
\fmf{photon,right=1}{vr1,b3}
\fmf{phantom,label=(d),l.dis=0.7cm}{b1,b3}
\end{fmfgraph*}}
&\\[2cm]
&
\displaystyle
\parbox[b]{20mm}{
\begin{fmfgraph*}(20,10)
\fmftopn{t}{3}
\fmfbottomn{b}{3}
\fmf{fermion}{b1,v1}
\fmf{fermion}{v3,b3}
\fmf{plain}{v2,v3}
\fmf{plain}{v1,v2}
\fmfv{d.sh=c,d.filled=empty,d.si=2mm}{b3}
\fmfv{d.sh=c,d.filled=empty,d.si=2mm}{b1}
\fmf{dots,tensio=.5}{t2,v2}
\fmffreeze
\fmf{photon,right=.5}{v1,v3}
\fmf{phantom,label=(e),l.dis=0.7cm}{b1,b3}
\end{fmfgraph*}
}
\qquad
 \parbox[b]{20mm}{
\begin{fmfgraph*}(20,10)
\fmftopn{t}{3}
\fmfbottomn{b}{3}
\fmf{fermion}{b1,v,b3}
\fmfv{d.sh=c,d.filled=empty,d.si=2mm}{b3}
\fmfv{d.sh=c,d.filled=empty,d.si=2mm}{b1}
\fmf{dots}{t2,v}
\fmffreeze
\fmf{phantom,label=(f),l.dis=0.7cm}{b1,b3}
\fmf{photon,right=.5}{b1,b3}
\end{fmfgraph*}
}
&
\\[5mm]
\end{eqnarray*}
\caption{One-loop diagrams contributing to the scattering of a charge.}
        \label{eb26b}
\end{figure}

\no Consider a charge scattered off a classical source (we denote
the vertex generically by $\Gamma$). At one loop the diagrams are those of
Fig.~\ref{eb26b}. Most of these diagrams are just propagator corrections
on one or other of the legs and as such have been effectively
calculated immediately above. The new diagrams are \ref{eb26b}e and
\ref{eb26b}f. We should point out that we here {\em only} display those
diagrams which can
generate an IR divergence and a pole for each of the external charged
legs; other figures see, e.g., Fig.'s~\ref{f5}i and \ref{f5}m,
will not yield poles in one of the legs\footnote{We should also point out
that even if the source is replaced by a photon coupling, as discussed
in the next section, then although there are more diagrams at one loop,
see Fig.~\ref{f5}, it is still the case that
only the diagrams of Fig.~\ref{eb26b} have the
correct pole structure.}.

For the covariant diagram,  \ref{eb26b}e,  we again take the $p\slsh+m$ numerator
factors through the interaction vertex $\gamma$ matrices and only retain
those terms which have a pole for each external leg. In this way we
obtain from this diagram
\begin{equation}
\Gamma^{ \ref{eb26b}\mathrm{e}}(p,p')= \frac {ie^2}{{p\slsh}'-m}
\Gamma \frac 1{p\slsh-m}
\intden D_{\rho\sigma}(k)\frac{{p'}^\rho}{p'\cdot k}
\frac{{p}^\sigma}{p\cdot k}
\,.
\end{equation}
Similarly, using Appendix \ref{fact_app}, the
one loop rainbow diagram, \ref{eb26b}f, may be factorised, and we so find
that its contribution to the vertex is
\begin{equation}
\Gamma^{ \ref{eb26b}\mathrm{f}}(p,p')= -\frac {ie^2}{{p\slsh}'-m}
\Gamma \frac 1{p\slsh-m}
\intden D_{\rho\sigma}(k)\frac{{V'}^\rho}{V'\cdot k}
\frac{{V}^\sigma}{V\cdot k}
\,.
\label{e-20.7c}
\end{equation}
We can now combine all the one loop vertex corrections from the
diagrams of Fig.\ \ref{eb26b} to find
\begin{eqnarray}
\Gamma^{{\mathrm{IR}}}(p,p')=
\frac {ie^2}{{p\slsh}'-m}\Gamma \frac 1{p\slsh-m}
&&\!\!\!\!\!\!\!\!\!\!\!
\intden
\Bigg\{\Bigg.
\left[  \frac{{p}^\rho}{p\cdot k}
-\frac{{V'}^\rho}{V'\cdot k}
\right]
D_{\rho\sigma}(k)
\left[  \frac{{V}^\sigma}{V\cdot k}
-\frac{{p'}^\sigma}{p'\cdot k}
\right]\nonumber \\
  &-&
\left[  \frac{{p}^\rho}{p\cdot k}
-\frac{{p'}^\rho}{p'\cdot k}
\right]
D_{\rho\sigma}(k)
\left[  \frac{{p}^\sigma}{p\cdot k}
-\frac{{p'}^\sigma}{p'\cdot k}
\right]
\Bigg.\Bigg\}
\,.
\end{eqnarray}
This result again makes the gauge invariant nature of our dressed Green's
functions manifest. Furthermore, the on-shell  equivalence between
$V^\mu/V\cdot k$ and $p^\mu/p\cdot k$ which we demonstrated above,
(\ref{mayodiez}),
implies that we can immediately  read off that,
when both the incoming and the outgoing charge are placed at
the correct points on the mass shell, the IR
divergences in the on-shell residue cancel.  (The spin independence of this
structure can again be easily seen by comparing with the results of
\cite{Bagan:1998kg}.) It should finally be noted
that the dressings associated with the two charges are different for
non-zero momentum transfer; there is then no gauge where
both dressings vanish and so our general dressed Green's functions
cannot be interpreted as ordinary Green's functions in some particular
gauge.

These cancellations are a strong indication that our dressings have the
physical significance which we argued in \I. However, as we will now show,
we can greatly improve on these one loop arguments.

\subsection{All Orders}
\label{AllO}

\no Having seen the cancellation of the IR divergences at leading order,
we now want to prove that this cancellation holds to all orders. To do
this we will show that the IR divergent
terms exponentiate; from this, together with the
one loop cancellation, the general statement follows.

\no \textbf{a) Propagating Charges}

\smallskip

\no A general proof of the IR finiteness of the usual matter
propagator in the Yennie and radiation gauges was provided many years ago by
Jackiw and Soloviev \cite{soloviev:1968} (for the non-covariant gauge
they had to use a particular mass-shell renormalisation point.)
Essentially these authors established that the (Lagrangian) matter
propagators are free of such
divergences if a function which they denote by $F_{R}$
vanishes (see their Eq.\ 3.41). This function is given by
\begin{equation}\label{ml162a}
F_{R}=\frac{e^2}{(2\pi)^3}\int\!d^4k \ee^{ik \cdot x}
\theta(k_{0})\delta(k^2)\frac{r_{\mu}r_{\nu}}{r^2}\Pi^{\mu\nu}
\,,
\end{equation}
where $r_{\mu}$ is the momentum of the matter, i.e., $m\gamma(\eta+v)$ in our
notation, and $\Pi^{\mu\nu}$ is $k^2$ times the photon propagator.
It is now easy to see that at the correct
point on the mass shell our matter propagators will be IR finite to any
order, since the dressed propagator is equivalent to the propagator of
the Lagrangian matter fields in that gauge where the dressing
vanishes. Now in such a dressing
gauge the photon
propagator is given by
\begin{equation}
D^{\mu\nu}=\frac1{k^2} \Pi^{\mu\nu}=\frac1{k^2}
\left[
-g^{\mu\nu}+\frac{V^{\mu}V^{\nu}}{V\cdot k}
-\frac{V^{2} k^\mu k^\nu}{(V\cdot k)^2}
\right]
\,,
\end{equation}
and so it follows that $u_\mu D^{\mu\nu}=0$ in the  region,
$k^2=0$, which is of course
enforced by the delta function in (\ref{ml162a}).
Therefore $F_{R}$ vanishes, and the dressed matter propagators are IR
finite to any order in perturbation theory,
if we renormalise at the appropriate
mass shell point, $p=m \gamma(\eta+v)$.

Although this shows that the IR divergences cancel, it is still
useful to study the general form of the
propagator so as to be able to investigate scattering.

\begin{figure}
$$
\begin{fmfgraph*}(40,22)
      \fmfleft{l}
      \fmfright{r}
      \fmf{heavy}{v1,v2}
      \fmf{brown_muck,tension=2}{v2,r}
      \fmf{brown_muck,tension=2}{l,v1}
      \fmffreeze
      \fmfv{d.sh=c,d.filled=empty,d.si=2mm}{l}
      \fmfv{d.sh=c,d.filled=empty,d.si=2mm}{r}
      \fmf{photon, right=0.87}{r,l}
      \fmf{photon, right=0.5,label=$\vdots$}{r,l}
  \end{fmfgraph*}
$$
        \caption{General diagram with rainbow lines
        for the dressed propagator.}
        \label{eb26a}
\end{figure}
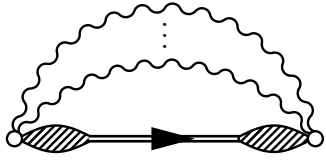
The general classes of diagrams which can yield IR divergences in the
on-shell propagator are shown in Fig.~\ref{eb26a}. The lines represent the
{\em full} electron and photon
propagators and the blobs correspond to all the diagrams involving
the free propagator or also
the set of all 1PI diagrams with contributions from the
dressing vertex, this we denote algebraically by  $G_v$ (see also (\ref{a2})
below).
Using
the factorisation property (see Appendix~\ref{fact_app}) we
immediately obtain for this figure
\begin{equation}\label{Cvvp}
G_v\frac i{p\slsh-m-\Sigma}G_v\; \ee^{-C_{vv}}\,,
\end{equation}
where $C_{vv}$ is defined by
\begin{equation}
C_{vv'}=-\intden \frac{V\cdot V'}{V\cdot k\, V'\cdot k
\, k^2}
\,.
\end{equation}
Again defining the renormalised
propagator as $\langle \psi^{B}\bar\psi^{B} \rangle=\Zmat v
\langle \psi^{R}\bar\psi^{R} \rangle \Zmatb v$,
we find that the wavefunction renormalisation constant has the
form
\begin{equation}
\Zmat v=\sqrt{Z_2^{\mathrm cov}} G_v e^{-C_{vv}/2}\,.
\label{10.5.99a}
\end{equation}
As discussed above, it follows from the work of Jackiw and Soloviev that
this is IR finite \emph{at the right point on the mass shell}.

\no \textbf{b) Scattering Charges}

\smallskip

\begin{figure}
$$
\begin{fmfgraph*}(40,22)
      \fmfleft{source} \fmfright{pone,ptwo}
      \fmf{dbl_dots,tension=3}{source,ver}
      \fmf{heavy}{ver,vv}
      \fmf{heavy}{v9,ver}
      \fmf{brown_muck,tension=2}{vv,pone}
      \fmf{brown_muck,tension=2}{v9,ptwo}
      \fmffreeze
      \fmfv{decor.shape=circle,decor.filled=shaded,decor.size=8mm}{ver}
      \fmfv{d.sh=c,d.filled=empty,d.si=2mm}{pone}
      \fmfv{d.sh=c,d.filled=empty,d.si=2mm}{ptwo}
      \fmf{photon, right=0.95}{pone,ptwo}
      \fmf{photon, right=0.18,label=$\stackrel{n}{\dots}$}{pone,ptwo}
  \end{fmfgraph*}
$$
        \caption{General diagram with rainbow lines
        for scattering off an external source (the dots).}
        \label{ml22b}
\end{figure}
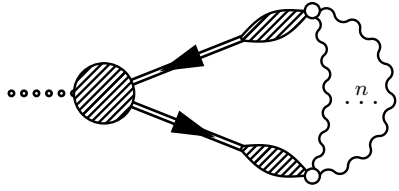
\no Consider again a charge scattered off a source. The diagrams
which can yield poles and IR divergences are shown in Fig.~\ref{ml22b}. Using
the factorisation property we so obtain
\begin{equation}
\ee^{-C_{vv'}}
G_v\frac i{p\slsh-m-\Sigma}V_{\mathrm{cov}}
\frac i{{{p\slsh}'}-m-\Sigma}
G_{v'}\,,
\end{equation}
where $V_{\mathrm{cov}}$ signifies the covariant (dressing
independent) central part of these diagrams. To obtain the $S$-matrix
element we multiply in with the inverse of the (IR finite)
renormalisation constants, (\ref{10.5.99a}),
on each side of the scattering vertex.
This kills off the $G_v$-type factors. We are
left with
\begin{equation}
\ee^{-C_{vv'}}\ee^{C_{vv}/2}\ee^{C_{v'v'}/2}\frac i{p\slsh-m-\Sigma}
V_{\mathrm{cov}} \frac i{{p\slsh}'-m-\Sigma}
\frac1{Z_{2}^{\mathrm{cov}}}
\,,
\end{equation}
which shows that the dressing contributions to the residues of the two
poles exponentiate. As it is well known (see, e.g., Chap.\ 13 of
\cite{weinberg:1995}) that the equivalent
dressing independent terms also exponentiate, it follows that all the
IR divergent terms in the residues are just the exponentiation of the
lowest order result. We have shown that these potential IR divergences
cancelled at one loop, hence it follows that this cancellation holds
to all orders.

The results of this section provide, we believe, extremely strong evidence
for the validity of our formalism. In \I\ we argued that any
description of charged  matter
needs to fulfill two requirements: \emph{gauge
invariance} and also the \emph{dressing equation}. We were able to solve
these conditions to obtain a description of dressed matter and argued that
the resulting structures could be identified with the soft and
phase structures which characterise the IR problem. In particular, we
demonstrated that the asymptotic limit of the fields
(\ref{theone})
corresponds to the particle modes (on the right point in the mass shell).
All the
calculations of this section indicate the correctness of these arguments
as we have shown that retaining the soft dressing suffices to remove the
soft divergences at all orders in perturbation theory from both the
on-shell Green's functions and from $S$-matrix elements where dressed
matter is scattered. Now we turn to the UV renormalisation of the
non-local and non-covariant fields which describe charged matter.



\section{Renormalisation}
\label{sec4}
\setlength{\unitlength}{1mm}

Higher loop effects will, as in any field theory, introduce
not only potential IR divergences but also UV singularities which we
have to regularise and renormalise. Recall that, to preserve gauge
invariance, we regulate the theory by dimensional regularisation.
This can after all be directly extended to non-abelian gauge theories.
Our notation for the (matrix) wave function
renormalisation constants was introduced in the
previous section. Here we shall compute operator insertions,
propagators and scattering and see where additional renormalisations
are required.

As far as our choice of renormalisation scheme is concerned, we will
use both on-shell and MS schemes. The latter may be most easily
generalised to QCD and since we will study composite operator
renormalisation, where there is not always a direct physical
interpretation, the MS scheme is the most natural. However, there is
no loss of generality since the UV divergent parts are the same in all
schemes and to obtain the $S$-matrix the finite parts in the various
Green's functions can be directly reinstated.

\subsection{One Insertion: The Dressed Electron Field as a Composite
Operator}
\label{ss1}
In this section we will investigate the properties of
the dressed electron, $\psi_{v}(x)$, understood as a
composite operator. In particular we will study its renormalisation at
the one-loop level. To this purpose, we shall
insert $\psi_{v}(x)$ in Green's functions with an arbitrary number of
photon fields. We shall show that these Green's functions can be
rendered UV finite by an appropriate multiplicative (matrix) renormalisation
of $\psi_{v}(x)$. We will demonstrate
that the dressed electron field does not mix under
renormalisation with other operators.
We would like to stress that, since the physical electron is
not a local operator, there is \textit{a
priori\/} no mathematical guarantee that it can be renormalised. It is
therefore highly significant to see that it
actually behaves just like a local operator
despite its being non-local and non-covariant.

\begin{figure}[tbp]
        \centering
         \parbox{20mm}{
\begin{fmfgraph*}(20,20)
\fmfleft{l}
\fmfright{r}
\fmf{fermion,label=(a),l.d=0.5cm}{l,r}
\fmfv{d.sh=c,d.filled=empty,d.si=2mm}{r}
\end{fmfgraph*}
}\qquad
+\quad
        \parbox{20mm}{
\begin{fmfgraph*}(20,20)
\fmfleft{l}
\fmfright{r}
\fmf{fermion}{l,v,r}
\fmffreeze
\fmf{photon,left=1}{v,r}
\fmfv{d.sh=c,d.filled=empty,d.si=2mm}{r}
\fmfv{d.sh=c,d.filled=empty,d.si=0mm,label=(b),l.d=0.5cm,l.a=-90}{v}
\end{fmfgraph*}
}
        \caption{Zero- and one-loop 1PI diagrams contributing to the
renormalisation of $\psi_{v}$.}
        \label{f1}
\end{figure}
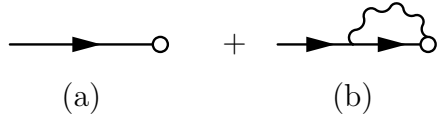

The Feynman rules for the different insertions of the dressing
have been obtained in Sect.~\ref{s1} and are summarised in Fig.~\ref{f3}.
Let us start by considering the simplest possible insertion
\begin{equation}
        \langle\psi_{v}(p)\bar\psi(p)\rangle\equiv
\int d^4x\;{\rm e}^{{i} p\cdot x} \langle0|{\rm
T}\psi_{v}(x)\bar\psi(0)|0\rangle\,.
        \label{a1}
\end{equation}
The corresponding one-particle irreducible (1PI) part, $G_{v}(p)$, is
defined as
\begin{equation}
        \langle\psi_{v}(p)\bar\psi(p)\rangle=G_{v}(p)\, {i}S(p)\,,
        \label{a2}
\end{equation}
where $S(p)$ is the standard full electron propagator. Up to one
loop, the diagrams contributing to $G_{v}(p)$ are those of Fig.~\ref{f1}.

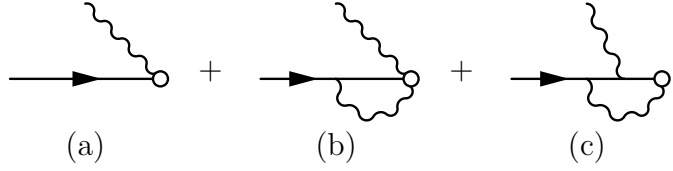
\begin{figure}[tbp]
\begin{eqnarray*}&
         \parbox[b]{20mm}{
\begin{fmfgraph*}(20,10)
\fmftopn{t}{3}
\fmfbottomn{b}{3}
\fmf{fermion}{b1,b3}
\fmfv{d.sh=c,d.filled=empty,d.si=2mm}{b3}
\fmf{photon}{t2,b3}
\fmfv{d.sh=c,d.filled=empty,d.si=0mm,label=(a),l.dis=0.7cm}{b2}
\end{fmfgraph*}
}\quad
+\quad
         \parbox[b]{20mm}{
\begin{fmfgraph*}(20,10)
\fmftopn{t}{3}
\fmfbottomn{b}{3}
\fmf{fermion}{b1,b2}
\fmf{plain}{b2,b3}
\fmfv{d.sh=c,d.filled=empty,d.si=2mm}{b3}
\fmf{photon}{t2,b3}
\fmffreeze
\fmf{photon,right=1}{b2,b3}
\fmfv{d.sh=c,d.filled=empty,d.si=0mm,label=(b),l.dis=0.7cm}{b2}
\end{fmfgraph*}
}\quad
+\quad
        \parbox[b]{20mm}{
\begin{fmfgraph*}(20,10)
\fmftopn{t}{3}
\fmfbottomn{b}{3}
\fmf{fermion}{b1,b2}
\fmf{plain,tension=0.3}{b2,v}
\fmf{plain,tension=0.3}{v,b3}
\fmfv{d.sh=c,d.filled=empty,d.si=2mm}{b3}
\fmffreeze
\fmf{photon}{t2,v}
\fmf{photon,right=1}{b2,b3}
\fmfv{d.sh=c,d.filled=empty,d.si=0mm,label=(c),l.dis=0.7cm}{b2}
\end{fmfgraph*}
}
&\\
\end{eqnarray*}
        \caption{Zero and one loop 1PI diagrams contributing to the
renormalisation of $\psi_{v}$.}
        \label{f2}
\end{figure}

One also needs to consider the insertion
\begin{equation}
        \langle\psi_{v}(p)\bar\psi(p-q)A^\mu(q)\rangle
        \,.
        \label{a3}
\end{equation}
The 1PI part of this, $e G_{v\,\rho}(p,q)$, is now defined through the
relation
\begin{equation}
        \langle\psi_{v}(p)\bar\psi(p-q)A^\mu(q)\rangle
        = \left[ G_{v}(p)\,{i} S(p)\,{i}e\Gamma_{\rho}(p,q)+
        e G_{v\,\rho}(p,q)\right]\,{i}S(p-q)\,{\rm
        i}D^{\mu\rho}(q)\,,
        \label{a4}
\end{equation}
where $D^{\mu\nu}(q)$ is the full photon propagator and
${i}e\Gamma_{\rho}(p,q)$ the full
vertex, i.e.,
\begin{equation}
{i}S(p)\,{i}e\Gamma_{\rho}(p,q){i}S(p-q)\,{
i}D^{\mu\rho}(q)
=\langle\psi(p)\bar\psi(p-q)A^{\mu}(q)\rangle\,.
        \label{b1}
\end{equation}
In Fig.~\ref{f2}
we have collected the corresponding 1PI diagrams up to one loop.

Likewise, one can define the 1PI portions of higher insertions.
Symbolically we write
\begin{equation}
e^n\,G_{v\,\rho_{1}\dots\rho_{n}}(p,q_{1},\dots,q_{n})=
\langle\psi_{v}(p)\bar\psi(p-{\scriptstyle \sum
}q_{i})A_{\rho_{1}}(q_{1})A_{\rho_{2}}(q_{2})
\dots A_{\rho_{n}}(q_{n})\rangle^{\rm 1PI}\,.
\label{b3}
\end{equation}
Eq.'s~(\ref{a2}) and~(\ref{a4}) are, of course, valid for both the bare
and renormalised Green's functions.

Let us now assume that $\psi_{v}(x)$
renormalises multiplicatively. More precisely, let us write
\begin{equation}
        \psi^{\rm B}_{v}(x)=\zmat v\psi^{\rm R}_{v}(x)\,.
        \label{b2}
\end{equation}
This assumption will be correct if we manage to prove that $\zmat v$
suffices to make \textit{all} $\psi_{v}(x)$ insertions, i.e., Eq.~\ref{b3},
UV finite.
It should be emphasised here that $\zmat v$ is not the renormalisation constant
introduced in~(\ref{Zmat}). $\zmat v$ renormalises Green's
functions with only one insertion of the dressed electron field,
whereas $\Zmat v$  is introduced  to deal with {\em two} insertions of the
dressed electron field. We will come back to this point
below.
From~(\ref{a2}) and~(\ref{a4}) it follows immediately
that
\begin{eqnarray}
        G^{\mathrm B}_{v}(p)&=&Z_{2}^{-1/2} \zmat v G^{\mathrm
        R}_{v}(p)\,,
                \label{b4}\\
        G^{\mathrm B}_{v\,\rho}(p,q)&=&Z_{2}^{-1/2}\zmat v G^{\mathrm
R}_{v\,\rho}(p,q)\,,
        \label{b4'}
\end{eqnarray}
where $Z_{2}$ is the usual electron wave function renormalisation
constant.
In obtaining~(\ref{b4'}) we have also used the QED identity
$Z_{e}=Z_{3}^{-1/2}$,
which amounts to charge universality:
\begin{equation}
        e^{\rm B} A^{\rm B}_{\mu}(x)=   e^{\rm R} A^{\rm R}_{\mu}(x)
\,.        \label{b5}
\end{equation}
Actually, the relation~(\ref{b4}) should hold for the 1PI portion of
\textit{any} $\psi_{v}(x)$ insertion.
\begin{equation}
G^{\rm B}_{v\,\rho_{1}\dots\rho_{n}}(p,q_{1},\dots,q_{n})
=Z_{2}^{-1/2}\zmat v G^{\rm
R}_{v\,\rho_{1}\dots\rho_{n}}(p,q_{1},\dots,q_{n})
        \,.\label{b8}
\end{equation}
This can be seen to be true by noticing that diagram~\ref{f2}c is UV
finite by
power counting. Therefore,
\begin{equation}
                G^{\rm B}_{v\,\rho}(p,q)=       G^{\rm B}_{v}(p)
{V_{\rho}\over V\cdot
                q}+\mbox{\sl UV finite}
        \,,\label{b6}
\end{equation}
which can be immediately generalised to
\begin{equation}
        G^{\rm B}_{v\,\rho_{1}\dots\rho_{n}}(p,q_{1},\dots,q_{n})=
        G^{\rm B}_{v}(p) {V_{1\,\rho_{1}}\over V_{1}\cdot
                q_{1}}\cdots
                {V_{n\,\rho_{n}}\over V_{n}\cdot
                q_{n}}+\mbox{\sl UV finite}\,.
        \label{b7}
\end{equation}
Thus, if $Z_{2}^{-1/2}\zmat v $ cancels the UV divergences in
Eq.~\ref{b4}, it is automatically guaranteed that $G^{\rm
R}_{v\,\rho}(p,q)$,
$G^{\rm R}_{v\,\rho_{1}\rho_{2}}(p,q_{1},q_{2})$, etc., defined
through~(\ref{b4'}) and~(\ref{b8}),
will also be UV finite.

The calculation of $G^{\rm B}_{v}(p)$ at one-loop (Fig.~\ref{f1}b)
is straightforward. From the Feynman rules we have
\begin{equation}
G^{\rm B\,(2)}_{v}(p)=
{i}e^2     \int{{\rm d}^{2\omega} k\over(2\pi)^{2\omega}}
{(/\kern-0.5em p-/\kern-0.5em k+m)\over [(p-k)^2-m^2]
 k^2}{/\kern-0.6em V\over V\cdot k}\,.
        \label{c1}
\end{equation}
The UV divergent part solely arises from
\begin{equation}
G^{\rm B\,(2)}_{v}(p)\Big|_{\rm div}=
-{i}e^2     \int{{\rm d}^{2\omega} k\over(2\pi)^{2\omega}}
{k^\alpha k^\beta\;[g_{\alpha\beta}- \gamma_{\alpha}(/\kern-0.5em\eta+
/\kern-0.5em v)(\eta-v)_{\beta} ]\over [(p-k)^2-m^2] k^2
[k^2-(k\cdot\eta)^2+
(k\cdot v)^2]}\,.
        \label{c2}
\end{equation}
With the help
of Appendix~\ref{ints_app} we obtain
\begin{equation}
G^{\rm B}_{v}(p)=1+{e^2\over16\pi^2}{1\over\hat\epsilon}+
        {e^2\over16\pi^2}{1\over\hat\epsilon}\left(
        {1\over\vb^2}+{1+\vb^2\over 2\vb^2}\chi(\vb)
        \right)/\kern-0.5em\eta
        /\kern-0.5em v +\mbox{\sl UV finite}
        \,.\label{c4}
\end{equation}
We recall now that in Feynman gauge
\begin{equation}
        Z_{2}=1-{e^2\over16\pi^2}{1\over\hat\epsilon}\,,
        \label{c5}
\end{equation}
and so we  finally find
\begin{equation}
        \zmat v =1+{e^2\over32\pi^2}{1\over\hat\epsilon}
        +
        {e^2\over16\pi^2}{1\over\hat\epsilon}\left(
        {1\over\vb^2}+{1+\vb^2\over 2\vb^2}\chi(\vb)
        \right)/\kern-0.5em\eta
        /\kern-0.5em v\,.
        \label{c6}
\end{equation}
Since we have found an explicit solution for Eq.~\ref{b2} we have
demonstrated that our assumption that these operators can be
multiplicatively renormalised (and do not mix with
other operators) is true. This is a very important result since it
would otherwise be a very complicated affair to define a dressing.
It is an open question whether this attractive property would hold
for gauge invariant dressed fields which do not obey our kinematical
requirements on the dressing (see Sect.\ 3 of \I).

\subsection{Two Insertions}

\subsubsection{The Dressed Propagator}

It is well known (see Chap.~6 of \cite{collins:1984}) that a Green's function
with several insertions of a renormalised composite operators requires in
general further renormalisation. The correlator of two electromagnetic currents
is a typical example. In that case (see Sect.~14.6 of \cite{collins:1984}), the lowest
order diagram is already at one-loop and has an overall divergence that cannot
be removed by multiplicative renormalisation. Instead, one has to take
derivatives to cancel the UV divergent parts. We shall see that the
situation~is quite different for the dressed electron field. First of all, the
lowest order diagram is just the standard electron propagator (no loops).
Hence, there is a possibility that multiplicative renormalisation works. We
shall show that this is the case and we shall recognise the extra
renormalisation constant to be that of the Isgur-Wise function or,
equivalently, that factor which is required to renormalise the cusp UV
divergences of Wilson lines. We shall then give an interpretation of this fact.

First, let us consider the correlator ${i}S_{v}(p)=\langle
\psi_{v}(p)\bar\psi_{v}(p)\rangle$. This is the dressed electron
propagator already discussed in~\cite{Bagan:1997su}. Here,
we shall recalculate
its UV divergent part and see where the different contributions to its
renormalisation constant come from. This will give us some insight into
the physics of charged matter. The relevant one-loop diagrams are shown
in Fig.~\ref{f4}.

Note that the diagrams~\ref{f4}a and~\ref{f4}b are essentially the same as
diagrams~\ref{f1}a and~\ref{f1}b
respectively, whereas~\ref{f4}c is the adjoint of~\ref{f2}b. All of
these diagrams have already been computed in Sect.~\ref{ss1}.
Diagram~\ref{f4}d
yields
\begin{equation}
{i}S^{\rm\protect\ref{f4}d}(p)={e^2\over/\kern-.5em p -m}
\int {{\rm
d}^{2\omega}k\over(2\pi)^{2\omega}}{V^2\over (V\cdot k)^2}
{/\kern-.5em k (/\kern-.5em p-/\kern-.5em k + m)\over[(p-k)^2-m^2]k^2}
\,.        \label{d5}
\end{equation}
As before $V^\mu=(\eta+v)^\mu(\eta-v)\cdot k-k^\mu$, and we have
made use of the identity
\begin{equation}
        {1\over /\kern-.5em p-/\kern-.5em k - m}
        ={1\over /\kern-.5em p - m}
        \left(1+/\kern-.5em k {1\over /\kern-.5em p-/\kern-.5em k - m}
\right)\,,
        \label{d6}
\end{equation}
and dropped massless tadpole integrals which vanish in dimensional
regularisation. The UV divergent part is
\begin{equation}
{i}S^{\rm\protect\ref{f4}d}(p)=
{-e^2\over/\kern-.5em p -m}     \int{{\rm d}^{2\omega}
k\over(2\pi)^{2\omega}}
        {k_{\mu}k_{\nu}
        \left[
        g^{\mu\nu}+\gamma^{-2}(\eta-v)^\mu(\eta-v)^\nu-2\eta^\mu \eta^\nu
        +2v^\mu v^\nu
        \right]
        \over [(p-k)^2-m^2][k^2-(k\cdot\eta)^2+(k\cdot v)^2]^2}
\,.        \label{d7}
\end{equation}
With the help of the integrals collected in Appendix~\ref{ints_app} we get
\begin{equation}
{i}S^{\rm\protect\ref{f4}d}(p)=
{{i}\over/\kern-.5em p -m}\;
{e^2\over8\pi^2}{1\over\hat\epsilon}\left[1+\chi(\vb)\right]
\,.        \label{d8}
\end{equation}
The full result for the UV-divergent part of the propagator up to one loop
is thus
\begin{eqnarray}
        {i}S^{\rm B}_{v}(p)&=&{{i}\over/\kern-.5em p -m}
        +G^{\rm B\;(2)}_{v}(p){{i}\over/\kern-.5em p -m}
        +{{i}\over/\kern-.5em p -m}\bar G^{\rm B\;(2)}_{v}(p)+
                {i}S^{\rm\protect\ref{f4}d\; B}_{v}(p)\nonumber\\
        &&+{{i}\over/\kern-.5em p -m}\left[-{i}\Sigma^{\rm
B\;(2)}(p)\right]
        {{i}\over/\kern-.5em p -m}
  \,,      \label{d9}
\end{eqnarray}
where $\Sigma^{\rm B\;(2)}(p)$ is the usual electron self-energy
\begin{equation}
        \Sigma^{\rm B\;(2)}(p)= {3 e^2\over
        16\pi^2}{1\over\hat\epsilon} m -
        { e^2\over
        16\pi^2}{1\over\hat\epsilon}(/\kern-.5em p -m)
        \,.\label{d12}
\end{equation}
Introducing the standard mass shift
\begin{equation}
        m\to m+\delta m,\qquad {\delta m\over m}= {3 e^2\over
        16\pi^2}{1\over\hat\epsilon}\,,
        \label{d10}
\end{equation}
Eq.~\ref{d9} can be written as
\begin{eqnarray}
                {i}S^{\rm B}_{v}(p)&=&\left\{1+
                \;{e^2\over16\pi^2}{1\over\hat\epsilon}
                \left[3+2\chi(\vb)\right]
                \right\}
                 {{i}\over/\kern-.5em p -m}\label{d11} \\
                &+& {e^2\over16\pi^2}{1\over\hat\epsilon}
                \left[{1\over\vb^2}+{1+\vb^2\over2\vb^2}\chi(\vb)\right]
                \left(
                /\kern-.5em \eta/\kern-.5em v
                {{i}\over/\kern-.5em p -m}-
                {{i}\over/\kern-.5em p -m}  /\kern-.5em
\eta/\kern-.5em v
                \right)\,.
        \nonumber
\end{eqnarray}
We immediately see that the field renormalisation $\zmat v$
introduced in~(\ref{b2}) and~(\ref{c6}), which would naively imply
the possible replacement
\begin{equation}
        S^{\rm B}_{v}(p){=}\zmat v S^{\rm R}_{v}(p) \zmatb v
       \,, \label{d13}
\end{equation}
cancels the non-covariant term
in~(\ref{d11}). However, an extra renormalisation is needed in order
to get rid of the first term and one must use the $\Zmat v$ renormalisation
matrix introduced in~(\ref{Zmat}). If we write $\Zmat v=Z_{v}^{1/2}\zmat v$,
we immediately find
\begin{equation}
        Z_{v}=1+{e^2\over8\pi^2}{1\over\hat\epsilon}
        \left[1+\chi(\vb^2)\right]\,,
        \label{d15}
\end{equation}
as  can be read off from Eq.~\ref{d8}.

In~\cite{Bagan:1997su}, we used the following ansatz for the multiplicative
renormalisation matrix, $\sqrt{Z_{2}}\exp(-{\rm
i}\sigma_{\mu\nu}\eta^\mu v^\nu\, Z'/Z_{2})$ for $\Zmat v$.
Our result in this paper is
\begin{equation}
          \Zmat v=1+
                \;{e^2\over32\pi^2}{1\over\hat\epsilon}
                \left[3+2\chi(\vb)\right]
                + {e^2\over16\pi^2}{1\over\hat\epsilon}
                \left[{1\over\vb^2}+{1+\vb^2\over2\vb^2}\chi(\vb)\right]
                /\kern-.5em \eta/\kern-.5em v
\,,     \label{e16.2a}
\end{equation}
which agrees with that of the previous calculation.
Note that $\Zmat v$ plays the role of $\sqrt{Z_{2}}$ in
standard QED.
It is the wave function renormalisation constant of the $\psi_{v}(x)$
field.

\subsubsection{The Photon-Dressed Electron Vertex}

The primary source of information on how electrons are scattered by
light is the basic correlator $\langle \psi_{v'}(p')\bar
\psi_{v}(p) A_\mu(q)
        \rangle$.
In this section we shall mainly be concerned
with its renormalisation.
For bare quantities, let us define the vertex,
$i e\,V^{\nu}_{vv'}(p,p')$, through
\begin{equation}
        i D^{\rm B}_{\mu\nu}(q)\;i e^{\rm B}\,V^{{\rm B}\,\nu}_{vv'}(p,p')=
        \langle \psi^{\rm B}_{v'}(p')\bar\psi^{\rm B}_{v}(p) A^{\rm B}_\mu(q)
        \rangle\,;\qquad q=p'-p
\,.        \label{d16}
\end{equation}
One should not
expect that the renormalisation constants introduced so far
suffice to make $V^\mu_{vv'}(p,p')$ UV finite. As we will see in a
moment, a new renormalisation constant, which depends on
the relative velocity, is required.
The renormalised version of~(\ref{d16}) becomes
\begin{equation}
        i D^{\rm R}_{\mu\nu}(q)\;i Z(v,v')
        e^{\rm R}\,V^{{\rm R}\,\nu}_{vv'}(p,p')=
        \langle \psi^{\rm R}_{v'}(p')\bar\psi^{\rm R}_{v}(p) A^{\rm R}_\mu(q)
        \rangle\,;\qquad q=p'-p
\,,        \label{d16'}
\end{equation}
where we have allowed for the above mentioned new renormalisation constant,
$Z(v,v')$. Note that combining Eqs.~(\ref{d16}) and~(\ref{d16'}) we
obtain the relation
\begin{equation}
        V^{{\rm B}\;\mu}_{vv'}(p,p')= Z(v,v')\Zmat{v'}\;
         V^{{\rm R}\;\mu}_{vv'}(p,p')\;\Zmatb{v}\,.
        \label{g6}
\end{equation}

The need for $Z(v,v')$ is not that surprising: correlators of
renormalised operators, such as that on the right hand side of~(\ref{d16'})
do not necessarily have to be UV finite. As shall be discussed
below, this, and its generalisation to higher Green's functions,
will be the last renormalisation constant that we shall
need in our approach. It will also become clear in a moment that there is a
good physical reason for the introduction of this renormalisation
constant.

The correlator $i e\,V^\mu_{vv'}(p,p')$
can be
related to the scattering of a physical electron by a classical
electromagnetic field.
From~(\ref{d16'}), it is also clear that
$Z(v,v') e^{\rm R}$ measures the coupling of the dressed
electron to a photon. Certainly, at the tree level we reproduce the
standard result, $i e\,V^{(0)\,\mu}_{vv'}(p,p')=i e\,\gamma^\mu$.
At zero momentum transfer, we expect that $Z(v,v) e^{\rm R}$
is the usual charge of the electron, since
our claim is that the dressed field $\psi_{v}$ actually describes the
experimentally observed electron. Thus, we expect
\begin{equation}
         Z(v,v)=1\,.
        \label{e-17.2c}
\end{equation}
at all
orders. This can actually be proved, as is shown in Appendix~\ref{appward}.
%
%
This is a very important result, since it ensures that the strength
with which photons couple to dressed electrons is given by the standard
electric charge $e$, as we claimed at the beginning of this section. One only needs to recall
that $e$ is defined at zero momentum transfer, i.e., in
the limit $q\to0$, $p'\to p$ and $v'\to v$.

\begin{figure}[tbp]
$$
\begin{array}{ccccccc}
&\quad&
         \parbox[b]{20mm}{
\begin{fmfgraph*}(20,10)
\fmftopn{t}{3}
\fmfbottomn{b}{3}
\fmf{fermion}{b1,v,b3}
\fmfv{d.sh=c,d.filled=empty,d.si=2mm}{b3}
\fmfv{d.sh=c,d.filled=empty,d.si=2mm}{b1}
\fmf{photon}{t2,v}
\fmffreeze
\fmf{phantom,label=(a),l.dis=0.7cm}{b1,b3}
\end{fmfgraph*}
}&\qquad\qquad&
         \parbox[b]{20mm}{
\begin{fmfgraph*}(20,10)
\fmftopn{t}{3}
\fmfbottomn{b}{3}
\fmf{fermion}{b1,b3}
\fmfv{d.sh=c,d.filled=empty,d.si=2mm}{b3}
\fmfv{d.sh=c,d.filled=empty,d.si=2mm}{b1}
\fmf{photon}{t2,b1}
\fmfv{d.sh=c,d.filled=empty,d.si=0mm,label=(b),l.dis=0.7cm}{b2}
\end{fmfgraph*}
}&\qquad&
         \parbox[b]{20mm}{
\begin{fmfgraph*}(20,10)
\fmftopn{t}{3}
\fmfbottomn{b}{3}
\fmf{fermion}{b1,b3}
\fmfv{d.sh=c,d.filled=empty,d.si=2mm}{b3}
\fmfv{d.sh=c,d.filled=empty,d.si=2mm}{b1}
\fmf{photon}{t2,b3}
\fmfv{d.sh=c,d.filled=empty,d.si=0mm,label=(c),l.dis=0.7cm}{b2}
\end{fmfgraph*}
}
\\[1.5cm]
        \parbox[b]{20mm}{
\begin{fmfgraph*}(20,10)
\fmftopn{t}{3}
\fmfbottomn{b}{3}
\fmf{phantom}{t2,vt1}
\fmf{phantom,tension=1/2}{vt1,vt2}
\fmf{phantom}{vt2,v}
\fmf{phantom}{b1,vl1}
\fmf{phantom,tension=1/2}{vl1,vl2}
\fmf{phantom}{vl2,v}
\fmf{phantom}{v,vr1}
\fmf{phantom,tension=1/2}{vr1,vr2}
\fmf{phantom}{vr2,b3}
\fmffreeze
\fmf{fermion}{vr1,vr2}
\fmf{fermion}{b1,v}
\fmf{plain}{v,vr1}
\fmf{plain}{vr2,b3}
\fmfv{d.sh=c,d.filled=empty,d.si=2mm}{b3}
\fmfv{d.sh=c,d.filled=empty,d.si=2mm}{b1}
\fmf{photon}{t2,v}
\fmf{photon,right=1}{vr1,vr2}
\fmf{phantom,label=(d),l.dis=0.7cm}{b1,b3}
\end{fmfgraph*}}
&&
        \parbox[b]{20mm}{
\begin{fmfgraph*}(20,10)
\fmftopn{t}{3}
\fmfbottomn{b}{3}
\fmf{phantom}{t2,vt1}
\fmf{phantom,tension=1/2}{vt1,vt2}
\fmf{phantom}{vt2,v}
\fmf{phantom}{b1,vl1}
\fmf{phantom,tension=1/2}{vl1,vl2}
\fmf{phantom}{vl2,v}
\fmf{phantom}{v,vr1}
\fmf{phantom,tension=1/2}{vr1,vr2}
\fmf{phantom}{vr2,b3}
\fmffreeze
\fmf{fermion}{vl1,vl2}
\fmf{fermion}{v,b3}
\fmf{plain}{b1,vl1}
\fmf{plain}{vl2,v}
\fmfv{d.sh=c,d.filled=empty,d.si=2mm}{b3}
\fmfv{d.sh=c,d.filled=empty,d.si=2mm}{b1}
\fmf{photon}{t2,v}
\fmf{photon,right=1}{vl1,vl2}
\fmf{phantom,label=(e),l.dis=0.7cm}{b1,b3}
\end{fmfgraph*}}
&&
         \parbox[b]{20mm}{
\begin{fmfgraph*}(20,10)
\fmftopn{t}{3}
\fmfbottomn{b}{3}
\fmf{fermion}{v1,v2}
\fmf{plain}{b1,v1}
\fmf{plain}{v2,b3}
\fmfv{d.sh=c,d.filled=empty,d.si=2mm}{b3}
\fmfv{d.sh=c,d.filled=empty,d.si=2mm}{b1}
\fmf{photon}{t2,b1}
\fmffreeze
\fmf{photon,right=1}{v1,v2}
\fmf{phantom,label=(f),l.dis=0.7cm}{b1,b3}
\end{fmfgraph*}
}&&
         \parbox[b]{20mm}{
\begin{fmfgraph*}(20,10)
\fmftopn{t}{3}
\fmfbottomn{b}{3}
\fmf{fermion}{v1,v2}
\fmf{plain}{b1,v1}
\fmf{plain}{v2,b3}
\fmfv{d.sh=c,d.filled=empty,d.si=2mm}{b3}
\fmfv{d.sh=c,d.filled=empty,d.si=2mm}{b1}
\fmf{photon}{t2,b3}
\fmffreeze
\fmf{photon,right=1}{v1,v2}
\fmf{phantom,label=(g),l.dis=0.7cm}{b1,b3}
\end{fmfgraph*}
}\\[1.5cm]
        \parbox[b]{20mm}{
\begin{fmfgraph*}(20,10)
\fmftopn{t}{3}
\fmfbottomn{b}{3}
\fmf{fermion}{b1,v1}
\fmf{fermion}{v3,b3}
\fmf{plain}{v2,v3}
\fmf{plain}{v1,v2}
\fmfv{d.sh=c,d.filled=empty,d.si=2mm}{b3}
\fmfv{d.sh=c,d.filled=empty,d.si=2mm}{b1}
\fmf{photon,tensio=.5}{t2,v2}
\fmffreeze
\fmf{photon,right=.5}{v1,v3}
\fmf{phantom,label=(h),l.dis=0.7cm}{b1,b3}
\end{fmfgraph*}
}&&&&&&\\[1.5cm]
        \parbox[b]{20mm}{
\begin{fmfgraph*}(20,10)
\fmftopn{t}{3}
\fmfbottomn{b}{3}
\fmf{phantom}{t2,vt1}
\fmf{phantom,tension=1/2}{vt1,vt2}
\fmf{phantom}{vt2,v}
\fmf{phantom}{b1,vl1}
\fmf{phantom,tension=1/2}{vl1,vl2}
\fmf{phantom}{vl2,v}
\fmf{phantom}{v,vr1}
\fmf{phantom,tension=1/2}{vr1,vr2}
\fmf{phantom}{vr2,b3}
\fmffreeze
\fmf{fermion}{b1,v}
\fmf{fermion}{vr1,b3}
\fmf{plain}{v,vr1}
\fmfv{d.sh=c,d.filled=empty,d.si=2mm}{b3}
\fmfv{d.sh=c,d.filled=empty,d.si=2mm}{b1}
\fmf{photon}{t2,v}
\fmf{photon,right=.5}{b1,vr1}
\fmf{phantom,label=(i),l.dis=0.7cm}{b1,b3}
\end{fmfgraph*}}
&&
        \parbox[b]{20mm}{
\begin{fmfgraph*}(20,10)
\fmftopn{t}{3}
\fmfbottomn{b}{3}
\fmf{phantom}{t2,vt1}
\fmf{phantom,tension=1/2}{vt1,vt2}
\fmf{phantom}{vt2,v}
\fmf{phantom}{b1,vl1}
\fmf{phantom,tension=1/2}{vl1,vl2}
\fmf{phantom}{vl2,v}
\fmf{phantom}{v,vr1}
\fmf{phantom,tension=1/2}{vr1,vr2}
\fmf{phantom}{vr2,b3}
\fmffreeze
\fmf{fermion}{v,b3}
\fmf{fermion}{b1,vl2}
\fmf{plain}{vl2,v}
\fmfv{d.sh=c,d.filled=empty,d.si=2mm}{b3}
\fmfv{d.sh=c,d.filled=empty,d.si=2mm}{b1}
\fmf{photon}{t2,v}
\fmf{photon,right=1}{b1,vl2}
\fmf{phantom,label=(j),l.dis=0.7cm}{b1,b3}
\end{fmfgraph*}}
&&
         \parbox[b]{20mm}{
\begin{fmfgraph*}(20,10)
\fmftopn{t}{3}
\fmfbottomn{b}{3}
\fmf{fermion}{b1,b2}
\fmf{plain}{b2,b3}
\fmfv{d.sh=c,d.filled=empty,d.si=2mm}{b3}
\fmfv{d.sh=c,d.filled=empty,d.si=2mm}{b1}
\fmf{photon}{t2,b1}
\fmffreeze
\fmf{photon,right=1}{b1,b2}
\fmf{phantom,label=(k),l.dis=0.7cm}{b1,b3}
\end{fmfgraph*}
}&&
         \parbox[b]{20mm}{
\begin{fmfgraph*}(20,10)
\fmftopn{t}{3}
\fmfbottomn{b}{3}
\fmf{fermion}{b1,b2}
\fmf{plain}{b2,b3}
\fmfv{d.sh=c,d.filled=empty,d.si=2mm}{b3}
\fmfv{d.sh=c,d.filled=empty,d.si=2mm}{b1}
\fmf{photon}{t2,b3}
\fmffreeze
\fmf{photon,right=1}{b1,b2}
\fmf{phantom,label=(l),l.dis=0.7cm}{b1,b3}
\end{fmfgraph*}
}\\[1.5cm]
\parbox[b]{20mm}{
\begin{fmfgraph*}(20,10)
\fmftopn{t}{3}
\fmfbottomn{b}{3}
\fmf{phantom}{t2,vt1}
\fmf{phantom,tension=1/2}{vt1,vt2}
\fmf{phantom}{vt2,v}
\fmf{phantom}{b1,vl1}
\fmf{phantom,tension=1/2}{vl1,vl2}
\fmf{phantom}{vl2,v}
\fmf{phantom}{v,vr1}
\fmf{phantom,tension=1/2}{vr1,vr2}
\fmf{phantom}{vr2,b3}
\fmffreeze
\fmf{fermion}{b1,vl2}
\fmf{fermion}{v,b3}
\fmf{plain}{vl2,v}
\fmfv{d.sh=c,d.filled=empty,d.si=2mm}{b3}
\fmfv{d.sh=c,d.filled=empty,d.si=2mm}{b1}
\fmf{photon}{t2,v}
\fmf{photon,right=.5}{vl2,b3}
\fmf{phantom,label=(m),l.dis=0.7cm}{b1,b3}
\end{fmfgraph*}}
&&
        \parbox[b]{20mm}{
\begin{fmfgraph*}(20,10)
\fmftopn{t}{3}
\fmfbottomn{b}{3}
\fmf{phantom}{t2,vt1}
\fmf{phantom,tension=1/2}{vt1,vt2}
\fmf{phantom}{vt2,v}
\fmf{phantom}{b1,vl1}
\fmf{phantom,tension=1/2}{vl1,vl2}
\fmf{phantom}{vl2,v}
\fmf{phantom}{v,vr1}
\fmf{phantom,tension=1/2}{vr1,vr2}
\fmf{phantom}{vr2,b3}
\fmffreeze
\fmf{fermion}{vr1,b3}
\fmf{fermion}{b1,v}
\fmf{plain}{v,vr1}
\fmfv{d.sh=c,d.filled=empty,d.si=2mm}{b3}
\fmfv{d.sh=c,d.filled=empty,d.si=2mm}{b1}
\fmf{photon}{t2,v}
\fmf{photon,right=1}{vr1,b3}
\fmf{phantom,label=(n),l.dis=0.7cm}{b1,b3}
\end{fmfgraph*}}
&&
         \parbox[b]{20mm}{
\begin{fmfgraph*}(20,10)
\fmftopn{t}{3}
\fmfbottomn{b}{3}
\fmf{fermion}{b1,b2}
\fmf{plain}{b2,b3}
\fmfv{d.sh=c,d.filled=empty,d.si=2mm}{b3}
\fmfv{d.sh=c,d.filled=empty,d.si=2mm}{b1}
\fmf{photon}{t2,b1}
\fmffreeze
\fmf{photon,right=1}{b2,b3}
\fmf{phantom,label=(o),l.dis=0.7cm}{b1,b3}
\end{fmfgraph*}
}&&
         \parbox[b]{20mm}{
\begin{fmfgraph*}(20,10)
\fmftopn{t}{3}
\fmfbottomn{b}{3}
\fmf{fermion}{b1,b2}
\fmf{plain}{b2,b3}
\fmfv{d.sh=c,d.filled=empty,d.si=2mm}{b3}
\fmfv{d.sh=c,d.filled=empty,d.si=2mm}{b1}
\fmf{photon}{t2,b3}
\fmffreeze
\fmf{photon,right=1}{b2,b3}
\fmf{phantom,label=(p),l.dis=0.7cm}{b1,b3}
\end{fmfgraph*}
}\\[1.5cm]
&&
 \parbox[b]{20mm}{
\begin{fmfgraph*}(20,10)
\fmftopn{t}{3}
\fmfbottomn{b}{3}
\fmf{fermion}{b1,v,b3}
\fmfv{d.sh=c,d.filled=empty,d.si=2mm}{b3}
\fmfv{d.sh=c,d.filled=empty,d.si=2mm}{b1}
\fmf{photon}{t2,v}
\fmffreeze
\fmf{phantom,label=(q),l.dis=0.7cm}{b1,b3}
\fmf{photon,right=.5}{b1,b3}
\end{fmfgraph*}
}
&&
         \parbox[b]{20mm}{
\begin{fmfgraph*}(20,10)
\fmftopn{t}{3}
\fmfbottomn{b}{3}
\fmf{fermion}{b1,b3}
\fmfv{d.sh=c,d.filled=empty,d.si=2mm}{b3}
\fmfv{d.sh=c,d.filled=empty,d.si=2mm}{b1}
\fmf{photon}{t2,b1}
\fmffreeze
\fmf{photon,right=.5}{b1,b3}
\fmf{phantom,label=(r),l.dis=0.7cm}{b1,b3}
\end{fmfgraph*}
}&&
         \parbox[b]{20mm}{
\begin{fmfgraph*}(20,10)
\fmftopn{t}{3}
\fmfbottomn{b}{3}
\fmf{fermion}{b1,b3}
\fmfv{d.sh=c,d.filled=empty,d.si=2mm}{b3}
\fmfv{d.sh=c,d.filled=empty,d.si=2mm}{b1}
\fmf{photon}{t2,b3}
\fmffreeze
\fmf{photon,right=.5}{b1,b3}
\fmf{phantom,label=(s),l.dis=0.7cm}{b1,b3}
\end{fmfgraph*}
}\\[1.5cm]
&&&&&&
\end{array}
$$
        \caption{Zero- and one-loop 1PI diagrams contributing to
$V^\mu_{vv'}(p,p')$.}
        \label{f5}
\end{figure}
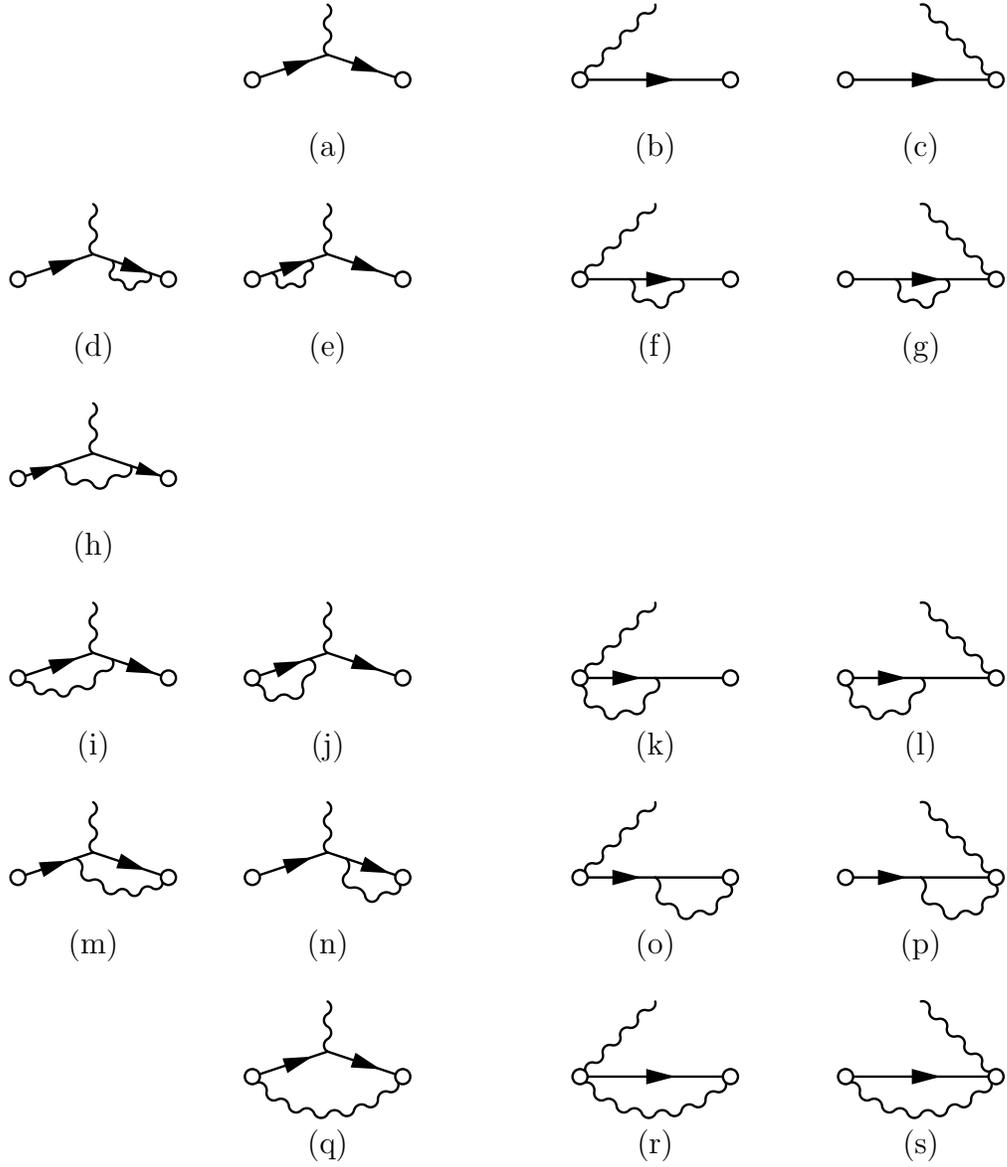

The relevant diagrams for the calculation of the vertex
are given in Fig.~\ref{f5}. Note
that, for simplicity, we do not include diagrams with vacuum
polarisation of the external photon line  since they are
effectively taken into account by using $e_{\rm R}$ instead of the bare
electric
charge $e_{\rm B}$ in the vertices. In the third and fourth
columns of  Fig.~\ref{f5}
we collect those diagrams for which the external photon is
directly attached to the dressing. For the diagrams in the first
two columns, the external photon goes into an interaction vertex.
Diagrams~\ref{f5}i and~\ref{f5}m are finite by power counting.
Diagrams~\ref{f5}a, \ref{f5}d, \ref{f5}e and~\ref{f5}h are standard QED
diagrams, i.e., they do not receive contributions from the dressings.
It is well known that~\ref{f5}d (or~\ref{f5}e) and~\ref{f5}h
cancel each other (once the electron mass has been
renormalised)
\begin{equation}
        -\left[{\rm Fig.~\ref{f5}h}\right]=\left[{\rm
Fig.~\ref{f5}d}\right]
        =\left[{\rm Fig.~\ref{f5}e}\right]=
        {{i}\over/\kern-.5em p'-m}\; 
        \gamma^\mu     \;
        {{i}\over/\kern-.5em
        p-m}\left(-{e^2\over16\pi^2}{1\over\hat\epsilon}\right)
        \label{g1}
        \,.
\end{equation}
This follows from the standard QED Ward identity $Z_{2}=Z_{1}$.
Hence, effectively, the first column can be dropped. All the other
diagrams can be easily computed from the results above
except for~\ref{f5}q, ~\ref{f5}r and~\ref{f5}s. The calculation of these
last three diagrams can be greatly simplified by choosing the Breit
frame where $\vb'=-\vb$. In this frame we have
\begin{equation}
\left[{\rm Fig.~\protect\ref{f5}r}\right]=
{{i}\over/\kern-.5em p' -m}{-eV^\mu\over V\cdot q}
e
\int{{\rm d}^{2\omega} k\over(2\pi)^{2\omega}}
        {k_{\nu}k_{\rho}
        \left[
        g^{\nu\rho}-\gamma^{-2}(\eta^\nu \eta^\rho
        -v^\nu v^\rho)
        \right]
        \over [(p'-k)^2-m^2][k^2-(k\cdot\eta)^2+(k\cdot v)^2]^2}
\,.        \label{g2}
\end{equation}
Using the results
of Appendix~\ref{ints_app} we finally obtain
\begin{equation}
\left[{\rm Fig.~\protect\ref{f5}r}\right]=
{{i}\over/\kern-.5em p' -m}{i\,V^\mu\over V\cdot q}
\;{e^2\over16\pi^2}{1\over\hat\epsilon}
(1-\vb^2)\chi(\vb)\,.
        \label{g3}
\end{equation}
Similarly
\begin{equation}
\left[{\rm Fig.~\protect\ref{f5}s}\right]=
{-i\,V'^\mu\over V'\cdot q}{{i}\over/\kern-.5em p -m}
\;{e^2\over16\pi^2}{1\over\hat\epsilon}
(1-\vb^2)\chi(\vb)
 \,,       \label{g4}
\end{equation}
where now $V'^\mu=(\eta+v')^\mu(\eta-v')\cdot
q-q^\mu=(\eta-v)^\mu(\eta+v)\cdot q-q^\mu$.

The calculation of diagram~\ref{f5}q might seem to be
much more involved. However,
since we are interested just in the UV
divergent part, which is easily seen to be
independent of the external momenta $p$,
$p'$ and $q$, we can set $p'=0$ to simplify the loop integral. We
end up with the same integral as in~(\ref{g2}). The final result is
\begin{equation}
\left[{\rm Fig.~\protect\ref{f5}q}\right]=
{{i}\over/\kern-.5em p' -m}\;\gamma^\mu\;
{{i}\over/\kern-.5em p -m}
\;{e^2\over16\pi^2}{1\over\hat\epsilon}
(1-\vb^2)\chi(\vb)
\,.        \label{g5}
\end{equation}
Note that the sum of diagrams~\ref{f5}q, \ref{f5}r and~\ref{f5}s is
just the
tree level result (diagrams~\ref{f5}a, \ref{f5}b and~\ref{f5}c) times
the UV divergent factor present in the three equations~(\ref{g3}), (\ref{g4})
and~(\ref{g5}).
Combining these with the results for the other diagrams we have
\begin{eqnarray}
        V^{{\rm B}\;\mu}_{vv'}(p,p')&=&V^{(0)\;\mu}_{vv'}(p,p')
\left\{1+{e^2\over16\pi^2}{1\over\hat\epsilon}(1-\vb^2)\chi(\vb)\right\}
        \nonumber\\
&+&
{e^2\over16\pi^2}{1\over\hat\epsilon}
                \left[{1\over\vb^2}+{1+\vb^2\over2\vb^2}\chi(\vb)\right]
\left(/\kern-.5em\eta/\kern-.5em v
V^{(0)\;\mu}_{vv'}(p,p')-V^{(0)\;\mu}_{vv'}(p,p')
        /\kern-.5em\eta/\kern-.5em v \right)
\,,        \label{i2}
\end{eqnarray}
where
$V^{(0)\;\mu}_{vv'}(p,p')$ is the tree level
correlator (the top row of Fig.~\ref{f5})
\begin{equation}
        V^{(0)\;\mu}_{vv'}(p,p')=
        {{i}\over/\kern-.5em p' -m}\;\gamma^\mu\;
{{i}\over/\kern-.5em p -m}+
{{i}\over/\kern-.5em p' -m}{i\,V^\mu\over V\cdot q}+
{-i\,V'^\mu\over V'\cdot q}{{i}\over/\kern-.5em p -m}
   \,.     \label{i1}
\end{equation}
From this we can read off that the non-covariant structures do
\emph{not} require any further renormalisation. Taking (\ref{g6})
and~(\ref{d15}) into account we find that the extra renormalisation
which is needed has the form
\begin{equation}
        Z(v,v')=1-{e^2\over16\pi^2}{1\over\hat\epsilon}
                \left[(1+\vb^2)\chi(\vb)+2\right]
        \,. \label{i3}
\end{equation}
To interpret this result, let us introduce $y=u\cdot u'$, where $u$  ($u'$)
is the four velocity associated with $\vb$ ($\vb'$). Then, in the Breit
frame
\begin{equation}
        y={1+\vb^2\over 1-\vb^2}\quad\Rightarrow\quad
        \log\left(y+\sqrt{y^2-1}\right)=-|\vb|\chi(\vb)
\,,        \label{i4}
\end{equation}
so that
\begin{equation}
                Z(v,v')=1+{e^2\over16\pi^2}{1\over\hat\epsilon}
                \left[{2y\over \sqrt{y^2-1}}
\log\left(y+\sqrt{y^2-1}\right)-2\right]
   \,.     \label{i5}
\end{equation}
The corresponding anomalous dimension is
\begin{equation}
        \gamma(v,v')=\mu{{\rm d}\over{\rm
d}\mu}\log{Z(v,v')}=-{e^2\over4\pi^2}
                \left[{2y\over \sqrt{y^2-1}}
\log\left(y+\sqrt{y^2-1}\right)-2\right]
        \,, \label{i6}
\end{equation}
which we recognise as the anomalous dimension of the Isgur-Wise
function~\cite{Isgur:1990ed,Isgur:1989vq}.
This is also the anomalous dimension of the cusp
divergences of Wilson loops. We will
return to the interpretation of this in a moment.

In the Breit frame, the limit of zero-momentum transfer corresponds
of course to $\vb\to 0$. It is easy to see that the term in the square
brackets in (\ref{i3}) vanishes in this limit and that our Ward
identity (\ref{e-17.2c}) is indeed fulfilled.

As a final check,
one can also compute $Z(v,v')$ in the rest frame of the incoming
particle, namely: $\vb\to0$, $\vb\to \vb'$.
In this case, the integral in~(\ref{g2}) becomes
\begin{equation}
\int{{\rm d}^{2\omega} k\over(2\pi)^{2\omega}}
        {k_{\nu}k_{\rho}
        \left[
        g^{\nu\rho}-\eta^\nu v^\rho-\eta^\nu \eta^\rho
        +v^\nu v^\rho)
        \right]
        \over [(p'-k)^2-m^2][k^2-(k\cdot\eta)^2]
        [k^2-(k\cdot\eta)^2+(k\cdot v)^2]}
        =-{1\over 8\pi^2}{1\over\hat\epsilon}
        \label{i7}\,,
\end{equation}
where we have used Appendix~\ref{ints_app}.
The final result is
\begin{equation}
        Z(v,v')=1-{e^2\over16\pi^2}{1\over\hat\epsilon}
                \left[2+\chi(\vb)\right] \,.
        \label{i8}
\end{equation}
Recalling that in this frame $y=1/\sqrt{1-\vb^2}$ we once more obtain~(\ref{i5}).

\no \textbf{Physical interpretation}

Smooth Wilson loops are known to be UV
finite
if the electric charge is renormalised in the usual way\footnote{There is
also a linear divergence proportional to the length of the loop
that can be absorbed by an appropriate renormalisation of the (infinite)
mass of the charged that it is driven along the loop. This UV
divergence does not show up in dimensional regularisation~\cite{Dotsenko:1980wb}.}.
If two smooth parts of a loop meet at a point to form a non-zero
angle $y$,
extra UV divergences arise~\cite{Brandt:1981kf}. These are the so called cusp
singularities or divergences.
They result from the photon exchange
between the two bit of the loop near the singular point or cusp.
Physically, the divergence signals the infinite
bremsstrahlung produced by the infinite acceleration at a cusp where
the heavy charge, which is being
driven along the loop, suffers a sudden change of
direction. It is well known that these UV cusp divergences require
further renormalisation. One needs to introduce a new renormalisation
constant which must depend on the angle of the cusp, $Z(y)$.

In the Heavy Quark Effective Theory (HQET) the same phenomenon takes place
when a heavy quark suddenly changes its velocity, e.g., after it is
\lq kicked\rq\ by a weak current. This is what happens in, for example,
the decay
$b\to c+\ell+\nu_{\ell}$, since, from the point of view of the HQET, the
two quarks $b$ and $c$ are identical; both are infinitely heavy
quarks; and it is just the quark¥s velocity that changes.
Heavy quark (meson) to heavy quark (meson) transitions are determined by a
universal form factor known as the Isgur-Wise function. This encodes
the dynamics of the light degrees of freedom that surrounds the heavy
quark in a heavy hadron. From our comments above, it follows that,
in  HQET, $Z(y)$ must renormalise the Isgur-Wise function. This is,
again, a standard result in HQET.

For our dressed charges, we are faced with the same situation. The dressing
provides a photonic cloud (just like the light degrees of freedom in
 HQET) which accompanies the charge in its journey to the interaction
region. This cloud is such that it provides the right
electromagnetic field for the charged particle: that of a charge moving with
constant (asymptotic) velocity, $\vb$. In the remote future, the
charges carries with it a cloud
appropriate to a different velocity, $\vb'$. From the point of view
of the photonic cloud the charge is infinitely heavy or classical
(the cloud cannot change the particle's velocity). As far as the
dressing is concerned, the scattering is merely an abrupt change of
velocity at the point where the two (straight) worldlines
characterised by $u=\gamma(1,\vb)$ and $u'=\gamma(1,\vb')$ intersect.
As a result the same infinite bremsstrahlung takes place and an extra
renormalisation constant $Z(y)$ is necessary.

\no \textbf{${\boldsymbol g}{\bf -2}$ for the dressed electron}

The next obvious one-loop test is the calculation of
the anomalous magnetic
moment of the dressed electron, i.e., $g-2$. We will show
that this quantity has the standard QED value, $e^2/4\pi^2$, thus reinforcing
our interpretation of the dressed electron as being the electron we
actually observe in the laboratory.
We need to compute the $S$-matrix element associated to
$V^\mu_{vv'}(p,p')$ and it is, therefore, convenient to work in the
on-shell renormalisation scheme. We recall that the relevant  diagrams
are in Fig.~\ref{f5}. Actually, only those in the first two columns
give a non-zero contribution to the $S$-matrix element, thus we
can write the $S$-matrix element as
\begin{equation}
\bar u(p,s'){ i}e\, \mathfrak M^{\rm R\mu}_{vv'}(p,p') u(p,s)
        \label{e-19.7a}
\end{equation}
where the electron momenta are on-shell, i.e., $p=m \ga (\eta+v)$, $p'=m \ga'
(\eta+v')$. Here $\mathfrak M^{\rm R\mu}_{vv'}(p,p')$ denotes
the renormalised diagrams that we referred to above after
a tree level propagator has been amputated from  each of the two fermion legs.
A more precise definition of~$\mathfrak M^{\rm R\mu}_{vv'}(p,p')$ can
be found in App.~\ref{appward}.
Note that diagrams~\ref{f5}i and~\ref{f5}m give no contribution
to~(\ref{e-19.7a})
because they have only a single pole in the external
electron momenta.

In the on-shell scheme, the self energy-like
diagrams~\ref{f5}d, \ref{f5}e, \ref{f5}j, \ref{f5}n, do {\em not} have to be
included since $\mathfrak M^{\rm R\mu}_{vv'}(p,p')$ in~(\ref{e-19.7a}) is a renormalised Green
function on the mass shell. This can be most easily seen by switching to the counterterm
language for a moment and then noticing that for each of these
diagrams there is a counterterm
diagram which exactly
cancels it on the mass shell. This is in particular true for these four diagrams.
The cancellation is complete because we
demand the renormalised full propagator of the dressed electron to be the
tree level one near the mass shell. One may still worry that this cancellation does not
necessarily have to take place diagram by diagram but only for their sum. This is actually
the case in general. However, for the argument we present below, it suffices to
show that the counterterm of diagrams~\ref{f5}j and~\ref{f5}n exactly
cancel against the corresponding loop diagrams, so that one can effectively drop them
completely in~(\ref{e-19.7a}). The counterterm of the loop in~\ref{f5}j
 is defined to be the value of the loop in diagram~\ref{f4}c
 but with the
opposite sign because none of the other
diagrams in Fig.~\ref{f4} are proportional to the structure
$[/\kern-0.5em p-m]^{-1} /\kern-0.5em\eta /\kern-0.5em v$
($/\kern-0.5em\eta /\kern-0.5em v'
[/\kern-0.5em p'-m]^{-1}$) --- the expression corresponding to
diagram~\ref{f4}d on the mass shell is
given in~(\ref{e-19.7b}) and is seen to be proportional to the tree level
fermion propagator\footnote{%
In dimensional regularisation this
diagram is zero because~(\protect\ref{e-19.7b}) is a massless
tadpole integral. This diagram is actually both IR and UV divergent
and the two $D-4$ poles cancel each other because IR and UV divergences
are treated in the same way in
dimensional regularisation. If we were regularizing IR divergences by
keeping the electron slightly off shell [see comments around
Eq.~(\protect\ref{e-20.7a})] then the integrand of~(\protect\ref{e-19.7b})
would have an extra factor
$$
{2p\cdot k-k^2\over 2p\cdot k - k^2+\Delta}\,.
$$
\label{fff}
},
as is that of diagram~\ref{f4}e.
By comparing~\ref{f5}j (\ref{f5}n) with~\ref{f4}c (\ref{f4}b) it is apparent
that the loop integrals are the same and hence the former diagram is
completely cancelled by its counterterm, as we claimed. (This argument can be directly
extended to diagram \ref{f5}n but now comparing with \ref{f4}b and making the
substitutions $v\to v'$ and $p\to p'$.)

From the previous paragraph it should now be clear that the only
non-standard contribution to $g-2$ can possibly come from
diagram~\ref{f5}q. It is, therefore, our task to compute the
coefficient $\delta F_{2}$ of the magnetic moment structure
$i\sigma^{\mu\nu}(p'-p)_{\nu}/(2m)$ for this diagram because $g-2$ is
given in terms of it by
\begin{equation}
    g-2={e^2\over 4\pi^2}+2\delta F_{2}
    \label{e-20.7b}
\end{equation}
The expression for this diagram has already been given in~(\ref{e-20.7c})
where\footnote{%
Again, in dimensional regularisation this integral is zero on the mass
shell (see footnote~\ref{fff}). If we chose to regularise the IR
divergences by introducing a small off-shellness $\Delta$, The
integrand of~(\protect\ref{e-20.7c}) would have an extra factor
$$
{2p'\cdot k-k^2\over 2p'\cdot k-k^2+\Delta}+
{2p\cdot k-k^2\over 2p\cdot k-k^2+\Delta}-
{2p'\cdot k-k^2\over 2p'\cdot k-k^2+\Delta}{2p\cdot k-k^2\over 2p\cdot k-k^2+\Delta}
$$
} one just has to substitute $\gamma_{\mu}$ for $\Gamma$. But then again we see
that this diagram is proportional to $\gamma_{\mu}$ (once the external
propagators have been amputated) and can contribute only to the form
factor widely known as $F_{1}$. Hence, $\de F_{2}=0$ and $g-2$ has its
standard QED value.


\section{Conclusions}

Physical variables in gauge theories have to fulfil Gauss' law,
i.e., they must  be locally gauge invariant. Since the coupling in
the presence of massless gauge bosons does \emph{not}
effectively \lq switch off\rq\ at very large times, Gauss' law
continues to impose a non-trivial constraint on the matter fields.
This directly implies that  the fundamental matter fields in
the original Lagrangian of QED should not be identified with the asymptotic
physical fields. In particular all charged fields whether quarks, gluons or
electrons must always carry with them a
chromo- or electromagnetic cloud and only these systems -- the matter
and its associated dressing cloud taken together -- can have
any physical meaning.

Of course there are many ways to produce such gauge invariant constructs
and the first task is to discover which descriptions have any physical
relevance. To this aim we required in \I\ two things of any description of
a physical particle: i) local gauge invariance; and ii) that the dressing
obeys an equation which depends upon the velocity of the charged particle.
In QED we were able to solve these requirements and so found a general
description of charged particles. These gauge invariant, dressed fields
were shown to have a precise physical interpretation in  the asymptotic
region: their modes, taken at the appropriate point on the mass
shell, are free particle modes. The distortion factor which
prevents any particle identification of the matter fields in
relativistic QED~\cite{kulish:1970} was demonstrated to be
cancelled by the dressing, which is strong evidence that our dressings
indeed incorporate the asymptotic dynamics.

In this paper we have further tested these
physical variables in a wide variety of field theoretical
calculations. Since any description of a physical charged particle
is necessarily both non-local and non-covariant, it is important to study
the renormalisability of these variables\footnote{We recall the notorious
renormalisation problems arising in axial gauges.}. Similarly we have
studied the infra-red problem which, as we argued in \I, results from the
non-trivial asymptotic dynamics which is not properly included in the Green's
functions of unphysical asymptotic fields. We have demonstrated in
this paper that our fields pass a series of such IR and UV tests
with flying colours.

In the soft domain we already knew \cite{Bagan:1997su,Bagan:1997dh}
that an IR finite, on-shell
renormalisation of the propagator of our fields was possible at one loop
(for both QED and scalar QED). This paper has presented efficient methods to
demonstrate this and, most importantly, shown that this
crucial property can be
extended to scattering. Since the dressing depends on the velocity of the
charge, in a scattering process we have to associate different dressings
to the charges in the distant past and far future. On-shell
renormalisation then corresponds to putting each of these fields on shell at
the various
appropriate points on the mass shell. We have seen that doing this in a
self-consistent manner leads to the cancellation of the IR divergences
which arise when we go on shell using unphysical asymptotic fields.
We stress that this is a very delicate cancellation and
that were we to go on shell at a point inappropriate to even one of the dressings
then the Green's functions would contain IR divergences and we would not be able
to construct $S$-matrix elements!

We were further able to show that the soft divergences,
in the on-shell Green's functions of our physical fields and the
corresponding $S$-matrix elements, cancel at all orders of perturbation theory.
We believe that these results are strong evidence  both of the validity
of our formalism and of the specific physical interpretation we attach to it.

We also note here in passing that, since for a particle being scattered the
appropriate dressings for the incoming and outgoing asymptotic
charges are different, there is no gauge such that both dressings in a
scattering process vanish (except for the special case of zero momentum
transfer when the dressings are of course identical).
This is in contrast to the case of the
propagator, where there is only one dressing which can be reduced to unity
by an appropriate gauge choice. There our formalism can give a
physical explanation for some results which were previously just curiosities,
such as the IR-finiteness of the on-shell electron propagator in Coulomb
gauge if renormalisation was performed at the static point of the mass
shell~\cite{Kibble:1968a}.

\bigskip

Since our charged fields are non-local and non-covariant it was not clear
in advance what their renormalisation properties would be. In
the one-loop propagator, we had previously demonstrated that
the UV divergences could be
multiplicatively renormalised in this Green's function (via a matrix
multiplication for the fermionic theory). This property is highly
non-trivial and we have here extended this to study the renormalisation of
our physical fields both considered as composite operators and also in vertex
functions.

We saw that under renormalisation
the dressed electron does \emph{not} mix with other operators. This
property was seen in the simplest possible insertion and is trivially the
case for higher Green's functions with one insertion of a physical field.

Proceeding to higher insertions, we considered the propagator and the
scattering vertex. We derived a Ward identity linking
the scattering vertex for dressed electrons with their propagation.
The important physical results which emerged from these studies were
that the dressed electron couples to the photon with
the usual coupling strength and that the anomalous magnetic moment
can be reproduced using physical fields.

\bigskip

There are many natural extensions of the work reported in these two
papers and we will conclude this paper by describing some of the more
obvious extensions.

Two clear areas where IR~structures directly impede phenomenological
progress are collinear divergences, which are a consequence  of the
masslessness of the (colour charged) gluons,
and  field theories at finite temperature where soft singularities
are much more severe.

The finite temperature case can be attacked in the framework of the abelian theory\footnote{It
is also worth noting that the lower dimensional abelian theory at zero temperature is
also plagued by severe soft divergences.} A first step in the struggle against collinear
divergences would be to consider QED with massless matter fields.
This is a natural testing ground for QCD,
see Sect.~5 of \cite{Horan:1998im} for an initial discussion of this problem.

Confinement makes it obvious that the asymptotic dynamics of QCD can not be neglected.
There are two immediate questions here: why do we not see coloured quarks and gluons as physical
states and, given this, why can we still describe hadrons so well in terms of
constituent quarks?

It is clear that the original fields of the QCD Lagrangian cannot be
directly identified with these constituents. The effective quarks of quark models
must rather be constructed from the Lagrangian fields. It may be
seen \cite{Lavelle:1997ty} that only
fields which are locally gauge invariant can have a well-defined colour
charge (which is required in any effective quark model).

The primary difference between QCD and QED is that
the vector bosons in non-abelian gauge theories carry colour charges and
directly interact amongst themselves. This raises a multiplicity of
questions, ranging from the structure of glueballs to the construction of
pomerons, which last it seems~\cite{Adloff:1997sc} is
dominated by glue and not quarks. It is, we
note, possible to construct gauge-invariant, dressed gluons in certain
domains. The possible phenomenological
importance for gauge invariant gluonic operators of dimension two has
been stressed many times.

 At lowest order in the QCD coupling
the anti-screening effect in the beta function
is twelve times as large as the opposed screening term.
The first non-trivial test of QCD dressings was a study of
the interquark potential presented in~\cite{Lavelle:1998dv}.
Here it was shown that the minimal
gauge-invariant extension to QCD
of the dressing appropriate to a static charge in
QED exactly describes the gluonic configuration which
is responsible for asymptotic freedom and anti-screening. This
showed that this anti-screening interaction takes place between
separately gauge-invariant constituents.
This is currently being
extended to higher orders in the coupling which will for the first
time clarify the interplay between anti-screening and screening
beyond the above mentioned lowest order effect.

The electron propagator discussed above is, at one loop,
the same as the quark propagator, apart from trivial colour factors.
Higher loop investigations will test the non-abelian nature of the
dressings appropriate for colour charges. A first test is clearly to use
the dressing of~\cite{Lavelle:1998dv}, generalised to arbitrary
velocities, in the two-loop
propagator. One would not
necessarily expect all the IR divergences to cancel (phase-type
structures, see \I, are missing in the minimal dressing) but, since the primary gluonic
effects are evidently included in the dressing, much should indeed
cancel. These studies
will open a new window on the interplay between asymptotic freedom and
collinear structures.

\bigskip\bigskip

\no\textbf{Acknowledgements:} This work was supported by the British
Council/Spanish Education Ministry \textit{Acciones Integradas} grant
no.\ 1801\thinspace /\thinspace
HB1997-0141. We thank Robin Horan, Tom Steele, Shogo Tanimura and Izumi Tsutsui
for discussions  and the organisers of the
XIXth UK Theory Institute where some of this work was carried out.
EB thanks the HEP group at BNL for their
warm hospitality and many interesting comments. He also acknowledges a
grant from the {\em Direcci\'on General de Ense\~nanza Superior e
Investigaci\'on Cient\'{\i}fica.}

\bigskip\bigskip\bigskip

\appendix


\appendix

\section{Factorisation of IR Divergences}\label{fact_app}

\no In this appendix we want to demonstrate that the IR divergences
associated with what we have called rainbow lines factorise to all orders
in perturbation theory. The proof will be diagrammatic.

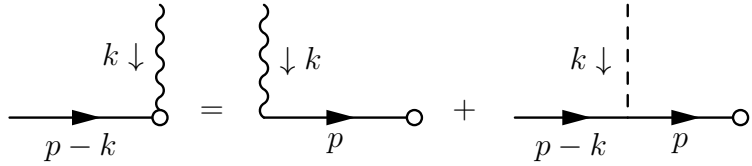
\begin{figure}
$$
\parbox[b]{20mm}{
\begin{fmfgraph*}(20,15)
\fmfstraight
\fmftopn{t}{3}
\fmfbottomn{b}{3}
\fmf{fermion,label=$p-k$}{b1,b3}
\fmf{photon,label=$k\downarrow$}{t3,b3}
 \fmfv{d.sh=c,d.filled=empty,d.si=2mm}{b3}
\end{fmfgraph*}
}\quad
=
\quad
\parbox[b]{20mm}{
\begin{fmfgraph*}(20,15)
\fmfstraight
\fmftopn{t}{3}
\fmfbottomn{b}{3}
\fmf{fermion,label=$p$}{b1,b3}
\fmf{photon,label=$\downarrow k$,l.s=left}{t1,b1}
 \fmfv{d.sh=c,d.filled=empty,d.si=2mm}{b3}
\end{fmfgraph*}
}
\quad
+
\quad
\parbox[b]{30mm}{
\begin{fmfgraph*}(30,15)
\fmfstraight
\fmftopn{t}{3}
\fmfbottomn{b}{3}
\fmf{fermion,label=$p-k$}{b1,b2}
\fmf{fermion,label=$p$}{b2,b3}
\fmf{dashes,label=$k\downarrow$}{t2,b2}
 \fmfv{d.sh=c,d.filled=empty,d.si=2mm}{b3}
\end{fmfgraph*}
}
$$
  \caption{Diagrammatic representation of the identity~(\protect{\ref{basic})}.}
        \label{ml22f}
\end{figure}

In a rainbow diagram it appears as though there are no poles
associated with the external legs which the rainbow lines are
attached to. We recall from Sect.~\ref{thr} that to extract
$S$-matrix elements, we need to find a pole for each particle entering
or leaving a scattering process. To be more precise, and again
recalling Sect.~\ref{thr},
we note that we are not looking for poles on a diagram by diagram
basis, and that they generally come accompanied by IR logarithms
of $m^2-p^2$ which will only cancel in the $S$-matrix elements or
in our dressed Green's functions when we go on shell at the right
point on the mass shell. However, we do need to extract explicit
factors of $1/(\not p-m)$ (which we
will henceforth for simplicity refer to as poles) from the individual
diagrams. To obtain these poles from the rainbow diagrams
we will below make much use of the general algebraic identities
\begin{eqnarray}\label{basic}
\frac1{\not p-\not k-m}&=&
\frac1{\not p-m}\left[1+\not k\frac1{\not p - \not k-m}\right]\,,
\\
 &=&
\left[1+\frac1{\not p - \not k-m}\not k\right]\frac1{\not p-m}\,.
\end{eqnarray}
The first line of (\ref{basic}) is represented in Fig.~\ref{ml22f}.
Although the vector boson lines are drawn explicitly, they are not
written in our identities since they are unaffected by our
manipulations. Note that we write the photon line in the second term as
a \emph{dashed} line to bring out the fact that there is now an extra
factor of $k$ in the numerator. This factor means that the
associated integrals over the loop momentum $k$ will now be
well defined (see below for examples) and in a diagram where all
rainbow lines have been replaced by dashed lines we can safely use
standard methods to read off the pole structure. We should stress
here that the dressing vertex factors are also unaffected by these
manipulations and that the circle should remind the reader that
these corrections are still present.

It is significant for what follows, since we want to study both the
on-shell IR behaviour of our Green's functions and the
UV~renormalisation properties, that the
terms of the form ${\not k}/({\not p}- {\not k} -m)$ reduce
to $-1$ both for on-shell $p$ and also in the UV limit of very large
$k$. Of course such identities are often used, e.g., to obtain
Ward identities.
To show the factorisation property we will proceed in this appendix by studying
first a one loop and then a two loop rainbow diagram. After these explicit
examples, we will consider the general case.
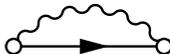
\begin{figure}[tbp]
        \centering
         \parbox{20mm}{
\begin{fmfgraph}(20,20)
\fmfleft{l}
\fmfright{r}
\fmf{fermion}{l,r}
\fmfv{d.sh=c,d.filled=empty,d.si=2mm}{r}
\fmfv{d.sh=c,d.filled=empty,d.si=2mm}{l}
\fmf{photon,left=0.5}{l,r}
\end{fmfgraph}
}
        \caption{The one-loop propagator rainbow diagram.}
        \label{appa1}
\end{figure}

\begin{figure}
    $$    \parbox[c]{20mm}{
\begin{fmfgraph*}(20,20)
\fmfleft{l}
\fmfright{r}
\fmf{phantom,tension=10}{l,l1}
\fmf{fermion,label=$\phantom{2}$}{l1,r}
\fmffreeze
\fmf{phantom,right=0.5}{l1,r}
\fmf{photon,right=0.5}{l1,l1}
\fmfv{d.sh=c,d.filled=empty,d.si=2mm}{r}
\fmfv{d.sh=c,d.filled=empty,d.si=2mm}{l1}
\end{fmfgraph*}
}\quad
+
\quad
        \parbox[c]{20mm}{
\begin{fmfgraph*}(20,20)
\fmfleft{l}
\fmfright{r}
\fmf{fermion,label=$\phantom{2}$}{l,v1}
\fmf{fermion}{v1,r}
\fmffreeze
\fmf{phantom,right=0.5}{l,r}
\fmf{dashes,left=1}{l,v1}
\fmfv{d.sh=c,d.filled=empty,d.si=2mm}{r}
\fmfv{d.sh=c,d.filled=empty,d.si=2mm}{l}
\end{fmfgraph*}}
$$
  \caption{The one-loop rainbow diagram re-expressed using
  Fig.~\protect\ref{ml22f}.}
        \label{ml22g}
\end{figure}
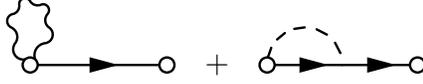

Consider now the simple one loop diagram, Fig.\ \ref{appa1}. This may be
written (up to irrelevant overall factors)
as\footnote{Note that although we work
in Feynman gauge in this appendix, the extension to a general gauge is
trivial.}
\begin{equation}
\intden \frac{V^2}{(V\cdot k)^2}\frac1{k^2}\frac1{[\not p-\not k-m]}
\,.
\end{equation}
One might be tempted to think that this diagram does not have a pole. However,
the non-local nature of the dressing vertices is such that we can indeed extract
a pole from this diagram as we now explain. We first note that the integrand is
ill-defined: for any $p$, i.e., even off shell, it diverges.
The factorisation property which we demonstrate here will allow
us to define such terms in a systematic way.
Using (\ref{basic}) the integrand may be written as a sum of two terms
\begin{equation}
\frac{V^2}{(V\cdot k)^2}\frac1{k^2}
\frac1{\not p-m}\Big[1+\not k\frac1{[\not p-\not k-m]}
\Big]
\,,
\end{equation}
the first integral is a massless tadpole and the second is
now well-defined. Diagrammatically this is shown in Fig.~\ref{ml22g}. We stress that
the tadpole and explicit pole structures are both clearly visible in the diagrams.

The second diagram has a UV divergence and, when $p$ is on-shell, it
also displays an IR divergence, i.e., a logarithm of $m^2-p^2$.
We now discard\footnote{We stress that the
infra-red divergence contained in this tadpole term is an off-shell divergence.
At no stage do we throw away IR divergences which first appear on shell.}
the massless tadpole since it vanishes in dimensional regularisation
and retain the second term which is, as advertised, IR finite off-shell.
Furthermore since on-shell (or in the UV region) the second factor
in (\ref{basic}) reduces to $-1$  we see that this loop \emph{in those regions}
does not receive any contribution from the fermion Feynman rules.

It is now apparent that, for our study of the IR structure,
this term is just the Feynman rule for the underlying diagram (the free
propagator)
times a subdiagram  which corresponds to the  integral $C_{vv}$
(recall Eq.\ \ref{Cvvp}) which
has been stripped off (factorised) from the initial diagram
(recall here Fig.\ \ref{appa2}). We thus see that, as we previously claimed,
we can extract a pole from a rainbow diagram.

To  factorise rainbow lines in more general diagrams, one merely
needs to make repeated use of
such manipulations. Let us further indicate how this works
with a two loop example (where we suppress the $1/2!$ symmetry factor).
\begin{equation}{\label{eb1b}}
        \parbox[c]{20mm}{
\begin{fmfgraph*}(20,20)
\fmfleft{l}
\fmfright{r}
\fmf{fermion,label=$\phantom{2}$}{l,r}
\fmffreeze
\fmf{photon,left=0.5}{l,r}
\fmf{photon,left=1.0}{l,r}
\fmfv{d.sh=c,d.filled=empty,d.si=2mm}{r}
\fmfv{d.sh=c,d.filled=empty,d.si=2mm}{l}
\end{fmfgraph*}
}
\;\approx\int\!\frac{d^4k_1}{{2\pi}^4}\frac{d^4k_2}{{2\pi}^4}
\frac{V^2_{k_1}}{(V_{k_1}\cdot
k_1)^2}\frac{V^2_{k_2}}{(V_{k_2}\cdot k_2)^2}\frac1{k_1^2k_2^2}
\frac1{\not p-\not k_1-\not
k_2-m}
\,.
\end{equation}
This diagram is ill-defined in both of the $k_i$
integrals. Using our basic identities we can write this as
\begin{eqnarray}
&&\nonumber\\
      &\displaystyle  \parbox[c]{20mm}{
\begin{fmfgraph*}(20,20)
\fmfleft{l}
\fmfright{r}
\fmf{phantom,tension=10}{l,l1}
\fmf{fermion}{l1,r}
\fmffreeze
\fmf{photon,left=0.5,label=$k_1$}{l1,r}
\fmf{photon,left=0.5,label=$k_2$}{l1,l1}
\fmfv{d.sh=c,d.filled=empty,d.si=2mm}{r}
\fmfv{d.sh=c,d.filled=empty,d.si=2mm}{l1}
\end{fmfgraph*}
}\quad
+
\quad
        \parbox[c]{20mm}{
\begin{fmfgraph*}(20,20)
\fmfleft{l}
\fmfright{r}
\fmf{fermion}{l,v1}
\fmf{fermion}{v1,r}
\fmffreeze
\fmf{photon,left=0.5,label=$k_1$}{l,r}
\fmf{dashes,right=1,label=$k_2$}{l,v1}
\fmfv{d.sh=c,d.filled=empty,d.si=2mm}{r}
\fmfv{d.sh=c,d.filled=empty,d.si=2mm}{l}
\end{fmfgraph*}
}\;
\approx\frac1{\not p-\not k_1-m}\left[1+\not k_2
\frac1{\not p-\not k_1-\not
k_2 -m}\right]\,.&\\[3mm]
&& \nonumber
\end{eqnarray}
where we have suppressed all factors other than the fermion propagator
since it alone is modified by the manipulations that show
factorisation.  The first term  vanishes as it
is a massless tadpole. The second structure is IR finite in $k_2$ off-shell
(but \emph{not} in $k_1$)
and does not yet have a pole in $1/(\not p-m)$. However,
we can see that
but for the remaining rainbow line it would have a pole. We therefore
again use (\ref{basic}) to rewrite this second term as
\begin{eqnarray}
&& \nonumber \\
        \parbox[c]{20mm}{
\begin{fmfgraph*}(20,20)
\fmfleft{l}
\fmfright{r}
\fmf{fermion}{l,v1}
\fmf{fermion}{v1,r}
\fmffreeze
\fmf{photon,left=1,label=$k_1$}{v1,r}
\fmf{dashes,right=1,label=$k_2$}{l,v1}
\fmfv{d.sh=c,d.filled=empty,d.si=2mm}{r}
\fmfv{d.sh=c,d.filled=empty,d.si=2mm}{l}
\end{fmfgraph*}
}
&+&
        \parbox[c]{20mm}{
\begin{fmfgraph*}(20,20)
\fmfleft{l}
\fmfright{r}
\fmf{plain}{l,v1}
\fmf{fermion}{v1,v2}
\fmf{plain}{v2,r}
\fmffreeze
\fmf{dashes,right=1,label=$k_2$}{l,v2}
\fmf{dashes,left=1,label=$k_1$}{v1,r}
\fmfv{d.sh=c,d.filled=empty,d.si=2mm}{r}
\fmfv{d.sh=c,d.filled=empty,d.si=2mm}{l}
\end{fmfgraph*}
}\;
\approx\frac1{\not p-\not k_1-m}\not k_2
\frac1{\not p-\not k_1-\not
k_2 -m}
\nonumber\\[3mm]
&\times&
\left[ 1+\not k_1\frac1{\not p-\not k_1-\not k_2-m}
\right]\frac1{\not p-\not k_2-m}\,.
\end{eqnarray}
We see that the second term here is well defined in both loops, but, as the diagram
makes apparent, it does not have a pole. (This can be seen algebraically as follows: if we try
to extract a pole in $1/(\not p-m)$ from
this figure using (\ref{basic}) the residue vanishes on shell.)

The first term, however, is still ill-defined in one of the
loops (the $k_1$ rainbow line is not yet a dashed line)
and may harbour a pole. Again employing
our  graphical transformations, Fig.~\ref{ml22f},   we can recast it as
\begin{eqnarray}
&&\nonumber \\
&\displaystyle        \parbox[c]{20mm}{
\begin{fmfgraph*}(20,20)
\fmfleft{l}
\fmfright{r}
\fmf{fermion}{l,v1}
\fmf{vanilla,tension=30}{v1,v2}
\fmf{vanilla,tension=30}{v2,v3}
\fmf{fermion}{v3,r}
\fmf{dashes,right=1,label=$k_2$}{l,v1}
\fmf{phantom,right=1}{v3,r}
\fmffreeze
\fmf{photon,tension=0.05,label=$k_1$}{v2,v2}
\fmfv{d.sh=c,d.filled=empty,d.si=2mm}{r}
\fmfv{d.sh=c,d.filled=empty,d.si=2mm}{l}
\end{fmfgraph*}
}
\;
+
\;
       \parbox[c]{20mm}{
\begin{fmfgraph*}(20,20)
\fmfleft{l}
\fmfright{r}
\fmf{plain}{l,v1}
\fmf{fermion}{v1,v2}
\fmf{plain}{v2,r}
\fmffreeze
\fmf{dashes,left=1,label=$k_1$}{v1,v2}
\fmf{dashes,right=1,label=$k_2$}{l,v1}
\fmfv{d.sh=c,d.filled=empty,d.si=2mm}{r}
\fmfv{d.sh=c,d.filled=empty,d.si=2mm}{l}
\end{fmfgraph*}
}\;
\approx\frac1{\not p-m}\left[1+\not k_1 \frac1{\not p-\not
k_1-m}\right]
\not k_2
\frac1{\not p-\not k_2-m}
\,,&
\end{eqnarray}
and the first diagram again vanishes since it is
a massless tadpole. The second diagram now displays a pole, $1/(\not p-m)$,
but the rainbow lines are now dashed and
the two internal loops reduce
on-shell to  $(-1)^2$. Reinstating the factors which we have suppressed we
obtain as expected the underlying diagram (here a free propagator) times a
factor of $(-C_{vv})^2/2!$. This example brings out all the fundamental ideas
of this method.

\begin{figure}
$$
\begin{fmfgraph}(25,20)
\begin{fmfsubgraph}(-.3w,0h)(.3w,1h)
\fmfcurved
\fmfleftn{b}{20}
\fmfrightn{a}{20}
\fmf{phantom}{a1,b1}
\fmf{phantom}{a2,b1}
\fmf{phantom}{a3,b1}
\fmf{phantom}{a20,b20}
\fmf{plain}{a4,b1}
\fmf{plain}{a5,b2}
\fmf{plain}{a6,b3}
\fmf{plain}{a7,b4}
\fmf{plain}{a8,b5}
\fmf{plain}{a9,b6}
\fmf{plain}{a10,b7}
\fmf{plain}{a11,b8}
\fmf{plain}{a12,b9}
\fmf{plain}{a13,b10}
\fmf{plain}{a14,b11}
\fmf{plain}{a15,b12}
\fmf{plain}{a16,b13}
\fmf{plain}{a17,b14}
\fmf{plain}{a18,b15}
\fmf{plain}{a19,b16}
\fmf{plain}{a20,b17}
\fmf{phantom}{a20,b18}
\fmf{phantom}{a20,b19}
\fmf{plain}{a4,a5,a6,a7,a8,a9,a10,a11,a12,a13,a14,a15,a16,a17,a18,a19,a20}
\end{fmfsubgraph}
\begin{fmfsubgraph}(.45w,.5h)(.6w,.6h)
\fmfstraight
\fmfbottomn{ll}{2}
\fmftopn{rr}{4}
\fmf{phantom}{ll1,ll2}
\fmf{photon}{ll2,vv1}
\fmf{phantom,tension=3}{vv1,rr1}
\fmf{photon}{rr4,ll2}
\fmfv{d.sh=c,d.filled=empty,d.si=2mm}{ll2}
\fmf{photon}{ll2,vv2}
\fmf{phantom,tension=8}{vv2,rr2}
\fmf{dots,left=.2}{vv2,rr4}
\end{fmfsubgraph}
\fmfcurved
\fmfleftn{l}{1}
\fmfrightn{r}{1}
\fmf{heavy,tension=.8}{l1,v1}
  \fmf{brown_muck}{v1,r1}
\end{fmfgraph}
$$
\caption{Part of a general diagram with $n$ rainbow lines emitted at an external
         charged leg.}
        \label{ml22a}
\end{figure}
Consider now a part of a general diagram, Fig.~\ref{ml22a},
where an external line is
connected to
one-particle irreducible factors involving the dressing at that line,
and to $n$
rainbow lines. These rainbow lines will render the diagram ill-defined.
We note that the matter
line is written as a double line to denote that the standard QED
counter\-term diagrams have been included so that the propagator has
a pole in $1/(\not p-m)$ with a logarithmic correction; this is
the full renormalised propagator of the Lagrangian matter field (the
photon propagator should also be understood as the full renormalised one).
This diagram could, for example, be a part of
the general scattering diagram of Fig.~\ref{ml22b} which, if it were
not for the presence of the rainbow lines, would
visibly have a pole in each leg.

We now proceed to take the end of one of the
rainbow lines to the left through the 1PI factor using the transformations of
Fig.~\ref{ml22f};
at every stage it will either move on to be attached at the
next vertex or it will
yield a well
defined integral. However, these well defined integrals will not have a
pole since
the line
which we are manipulating will extend from an external vertex into the
1PI structure. Therefore, since we only wish to find the poles
and their residues,
we drop these terms and obtain eventually Fig.~\ref{ml22c}.
\begin{figure}
$$
\begin{fmfgraph}(25,20)
\begin{fmfsubgraph}(-.3w,0h)(.3w,1h)
\fmfleftn{b}{20}
\fmfrightn{a}{20}
\fmf{phantom}{a1,b1}
\fmf{phantom}{a2,b1}
\fmf{phantom}{a3,b1}
\fmf{phantom}{a20,b20}
\fmf{plain}{a4,b1}
\fmf{plain}{a5,b2}
\fmf{plain}{a6,b3}
\fmf{plain}{a7,b4}
\fmf{plain}{a8,b5}
\fmf{plain}{a9,b6}
\fmf{plain}{a10,b7}
\fmf{plain}{a11,b8}
\fmf{plain}{a12,b9}
\fmf{plain}{a13,b10}
\fmf{plain}{a14,b11}
\fmf{plain}{a15,b12}
\fmf{plain}{a16,b13}
\fmf{plain}{a17,b14}
\fmf{plain}{a18,b15}
\fmf{plain}{a19,b16}
\fmf{plain}{a20,b17}
\fmf{phantom}{a20,b18}
\fmf{phantom}{a20,b19}
\fmf{plain}{a4,a5,a6,a7,a8,a9,a10,a11,a12,a13,a14,a15,a16,a17,a18,a19,a20}
\end{fmfsubgraph}
\begin{fmfsubgraph}(.45w,.5h)(.6w,.6h)
\fmfstraight
\fmfbottomn{ll}{2}
\fmftopn{rr}{4}
\fmf{phantom}{ll1,ll2}
\fmf{phantom}{ll2,vv1}
\fmf{phantom,tension=3}{vv1,rr1}
\fmf{photon}{rr4,ll2}
\fmfv{d.sh=c,d.filled=empty,d.si=2mm}{ll2}
\fmf{photon}{ll2,vv2}
\fmf{phantom,tension=8}{vv2,rr2}
\fmf{dots,left=.2}{vv2,rr4}
\end{fmfsubgraph}
\fmfcurved
\fmfleftn{l}{1}
\fmfrightn{r}{1}
\fmf{heavy,tension=.8}{l1,v1}
  \fmf{brown_muck}{v1,r1}
  \fmffreeze
\fmf{photon}{v1,md}
\fmf{phantom,tension=2}{md,rr1}
\end{fmfgraph}
$$
        \caption{The diagram of Fig.~\protect{\ref{ml22a}} after taking one end of the
        rainbow line through the visible 1PI blob.}
        \label{ml22c}
\end{figure}
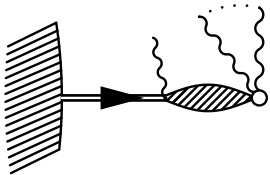

\begin{figure}
$$
\parbox{25mm}{
\begin{fmfgraph*}(25,20)
\begin{fmfsubgraph}(-.3w,0h)(.3w,1h)
\fmfleftn{b}{20}
\fmfrightn{a}{20}
\fmf{phantom}{a1,b1}
\fmf{phantom}{a2,b1}
\fmf{phantom}{a3,b1}
\fmf{phantom}{a20,b20}
\fmf{plain}{a4,b1}
\fmf{plain}{a5,b2}
\fmf{plain}{a6,b3}
\fmf{plain}{a7,b4}
\fmf{plain}{a8,b5}
\fmf{plain}{a9,b6}
\fmf{plain}{a10,b7}
\fmf{plain}{a11,b8}
\fmf{plain}{a12,b9}
\fmf{plain}{a13,b10}
\fmf{plain}{a14,b11}
\fmf{plain}{a15,b12}
\fmf{plain}{a16,b13}
\fmf{plain}{a17,b14}
\fmf{plain}{a18,b15}
\fmf{plain}{a19,b16}
\fmf{plain}{a20,b17}
\fmf{phantom}{a20,b18}
\fmf{phantom}{a20,b19}
\fmf{plain}{a4,a5,a6,a7,a8,a9,a10,a11,a12,a13,a14,a15,a16,a17,a18,a19,a20}
\end{fmfsubgraph}
\begin{fmfsubgraph}(.45w,.5h)(.6w,.6h)
\fmfstraight
\fmfbottomn{ll}{2}
\fmftopn{rr}{4}
\fmf{phantom}{ll1,ll2}
\fmf{phantom}{ll2,vv1}
\fmf{phantom,tension=3}{vv1,rr1}
\fmf{photon,label=$\phantom{2}$}{rr4,ll2}
\fmfv{d.sh=c,d.filled=empty,d.si=2mm}{ll2}
\fmf{photon}{ll2,vv2}
\fmf{phantom,tension=8}{vv2,rr2}
\fmf{dots,left=.2}{vv2,rr4}
\end{fmfsubgraph}
\fmfcurved
\fmfleftn{l}{1}
\fmfrightn{r}{1}
\fmf{heavy,tension=.8}{l1,v1}
  \fmf{brown_muck}{v1,r1}
  \fmffreeze
\fmf{photon}{v1,v1}
\end{fmfgraph*}
}
\quad+\kern10mm
\parbox{25mm}{
\begin{fmfgraph*}(25,20)
\begin{fmfsubgraph}(-.3w,0h)(.3w,1h)
\fmfleftn{b}{20}
\fmfrightn{a}{20}
\fmf{phantom}{a1,b1}
\fmf{phantom}{a2,b1}
\fmf{phantom}{a3,b1}
\fmf{phantom}{a20,b20}
\fmf{plain}{a4,b1}
\fmf{plain}{a5,b2}
\fmf{plain}{a6,b3}
\fmf{plain}{a7,b4}
\fmf{plain}{a8,b5}
\fmf{plain}{a9,b6}
\fmf{plain}{a10,b7}
\fmf{plain}{a11,b8}
\fmf{plain}{a12,b9}
\fmf{plain}{a13,b10}
\fmf{plain}{a14,b11}
\fmf{plain}{a15,b12}
\fmf{plain}{a16,b13}
\fmf{plain}{a17,b14}
\fmf{plain}{a18,b15}
\fmf{plain}{a19,b16}
\fmf{plain}{a20,b17}
\fmf{phantom}{a20,b18}
\fmf{phantom}{a20,b19}
\fmf{plain}{a4,a5,a6,a7,a8,a9,a10,a11,a12,a13,a14,a15,a16,a17,a18,a19,a20}
\end{fmfsubgraph}
\begin{fmfsubgraph}(.45w,.5h)(.6w,.6h)
\fmfstraight
\fmfbottomn{ll}{2}
\fmftopn{rr}{4}
\fmf{phantom}{ll1,ll2}
\fmf{phantom}{ll2,vv1}
\fmf{phantom,tension=3}{vv1,rr1}
\fmf{photon,label=$\phantom{2}$}{rr4,ll2}
\fmfv{d.sh=c,d.filled=empty,d.si=2mm}{ll2}
\fmf{photon}{ll2,vv2}
\fmf{phantom,tension=8}{vv2,rr2}
\fmf{dots,left=.2}{vv2,rr4}
\end{fmfsubgraph}
\fmfcurved
\fmfleftn{l}{1}
\fmfrightn{r}{1}
\fmf{heavy,tension=1.3}{l1,zz}
\fmf{dbl_plain,tension=5}{zz,md}
\fmf{dbl_plain}{md,v1}
  \fmf{brown_muck}{v1,r1}
  \fmffreeze
\fmf{dashes,right}{v1,md}
\end{fmfgraph*}
}
$$
  \caption{The previous figure when the other end of the rainbow line has been
  brought through the rest of the diagram.}
        \label{ml22d}
\end{figure}
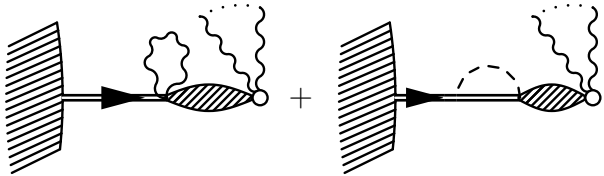

\begin{figure}
$$
\begin{fmfgraph*}(40,20)
\begin{fmfsubgraph}(-.2w,0h)(.2w,1h)
\fmfleftn{b}{20}
\fmfrightn{a}{20}
\fmf{phantom}{a1,b1}
\fmf{phantom}{a2,b1}
\fmf{phantom}{a3,b1}
\fmf{phantom}{a20,b20}
\fmf{plain}{a4,b1}
\fmf{plain}{a5,b2}
\fmf{plain}{a6,b3}
\fmf{plain}{a7,b4}
\fmf{plain}{a8,b5}
\fmf{plain}{a9,b6}
\fmf{plain}{a10,b7}
\fmf{plain}{a11,b8}
\fmf{plain}{a12,b9}
\fmf{plain}{a13,b10}
\fmf{plain}{a14,b11}
\fmf{plain}{a15,b12}
\fmf{plain}{a16,b13}
\fmf{plain}{a17,b14}
\fmf{plain}{a18,b15}
\fmf{plain}{a19,b16}
\fmf{plain}{a20,b17}
\fmf{phantom}{a20,b18}
\fmf{phantom}{a20,b19}
\fmf{plain}{a4,a5,a6,a7,a8,a9,a10,a11,a12,a13,a14,a15,a16,a17,a18,a19,a20}
\end{fmfsubgraph}
\fmfcurved
\fmfleftn{l}{1}
\fmfrightn{r}{1}
\fmf{heavy,tension=1.3}{l1,zz}
\fmf{dbl_plain,tension=5}{zz,md1}
\fmf{dbl_plain}{md1,md2}
\fmf{dbl_plain,label=$\displaystyle\underbrace{\kern2.3cm}_{n}$}{md2,md3}
\fmf{dbl_dashes}{md3,v1}
  \fmf{brown_muck}{v1,r1}
\fmfv{d.sh=c,d.filled=empty,d.si=2mm}{r1}
  \fmffreeze
\fmf{dashes,left}{md1,md2}
\fmf{dashes,left}{md2,md3}
\end{fmfgraph*}
$$
  \caption{The final result for the diagram after factorising the $n$ rainbow lines.}
        \label{ml22e}
\end{figure}
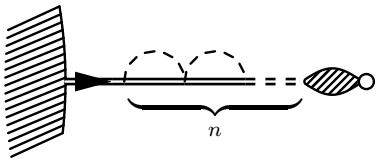

We now move to the other end of the line which we have been manipulating
and again use the identities to bring it towards the right.
Once again  the diagram
which yields a well defined integral will not contain a pole in the
matter line which was drawn in Fig.~\ref{ml22a}, and may be dropped. We
repeat this procedure until the second end of the rainbow line
reaches the end which we initially manipulated. We then have the
diagrams of Fig.~\ref{ml22d}. The massless tadpole vanishes but we see that
the remaining diagram has a pole -- up to the as yet unmanipulated
rainbow lines. Repeating this
procedure for all the rainbow lines  will yield Fig.~\ref{ml22e}.
Going on shell each of these loops can be removed and factorisation yields
$n$~powers of the form $-C_{vv'}$.

Finally we note that if a rainbow line connects two different
fermion lines, as might be the case in, say, M{\o}ller scattering,
then we have to supplement (\ref{basic}) by its photonic
equivalent
\begin{equation}
\frac1{(q-k)^2}=\frac1{q^2}\left[
1+\frac{2q\cdot k-k^2}{(q-k)^2}
\right]\,,
\end{equation}
which may be expressed diagrammatically in a fashion equivalent to
Fig.~\ref{ml22f} with the fermion replaced by a hard photon propagator. With
this result the above proof may be generalised to any process.



\section{Useful Integrals}\label{ints_app}


Here we collect some integrals
\begin{eqnarray}
        \int{{\rm d}^{2\omega} k\over(2\pi)^{2\omega}}
        {k_{\mu}k_{\nu}\over [(p-k)^2-m^2]k^2[k^2-(k\cdot\eta)^2+(k\cdot
v)^2]}
        &=&{{ i}\over 32\pi^2}{1\over\hat\epsilon}
        \left[A \eta_{\mu}\eta_{\nu} \right.    \label{e1}\\
        &+& \left. B (\eta_{\mu}\eta_{\nu}-g_{\mu\nu}) +
        C v_{\mu}v_{\nu}\right] \,,
\nonumber
\end{eqnarray}
where
\begin{eqnarray}
        A & = & -\chi(\vb) \,,
        \label{e2}  \\
        B & = & -{1\over\vb^2}-{1-\vb^2\over2\vb^2}\chi(\vb)\,,
        \label{e3}  \\
        C & = & {3\over \vb^4}+{3-\vb^2\over2\vb^4}\chi(\vb)\,.
        \label{e4}
\end{eqnarray}
Here we have defined
\begin{equation}
\chi(\vb)=\frac1{\vert\vb\vert}\log
\frac{1+\vert\vb\vert}{1-\vert\vb\vert}\,.
\end{equation}
We also need the integrals
\begin{eqnarray}
        \int{{\rm d}^{2\omega} k\over(2\pi)^{2\omega}}
        {k_{\mu}k_{\nu}\over [(p-k)^2-m^2][k^2-(k\cdot\eta)^2+(k\cdot
v)^2]^2}
        &=&{{ i}\over 32\pi^2}{1\over\hat\epsilon}
        \left[A \eta_{\mu}\eta_{\nu} \right.    \label{d1}\\
        &+& \left. B (\eta_{\mu}\eta_{\nu}-g_{\mu\nu}) +
        C v_{\mu}v_{\nu}\right] \,,
\nonumber
\end{eqnarray}
where
\begin{eqnarray}
        A & = & -{2\over1-\vb^2}+\chi(\vb)\,,
        \label{d2}  \\
        B & = & {1\over\vb^2}+{1+\vb^2\over2\vb^2}\chi(\vb)\,,
        \label{d3}  \\
        C & = & -{3-\vb^2\over
\vb^4(1-\vb^2)}-{3+\vb^2\over2\vb^4}\chi(\vb)\,.
        \label{d4}
\end{eqnarray}
Finally
\begin{eqnarray}
        \int{{\rm d}^{2\omega} k\over(2\pi)^{2\omega}}
        &&
        {k_{\mu}k_{\nu}\over
[(p-k)^2-m^2][k^2-(k\cdot\eta)^2][k^2-(k\cdot\eta)^2+
        (k\cdot v)^2]}
        \\
        && \qquad\qquad \qquad
        ={{i}\over 32\pi^2}{1\over\hat\epsilon}
        \left[A \eta_{\mu}\eta_{\nu} \right.    \label{h4}
       + \left. B (\eta_{\mu}\eta_{\nu}-g_{\mu\nu}) +
        C v_{\mu}v_{\nu}\right] \,,
\nonumber
\end{eqnarray}
where
\begin{eqnarray}
        A & = & 2\chi(\vb) \,,
        \label{h1}  \\
        B & = & -{2\over\vb^2}-{1-\vb^2\over\vb^2}\chi(\vb)
        \,, \label{h2}  \\
        C & = & {6\over \vb^4}+{3-\vb^2\over\vb^4}\chi(\vb)
        \,. \label{h3}
\end{eqnarray}
This result can be easily obtained by writing~(\ref{h4}) as
\begin{equation}
        \int_{0}^1 {\rm d} x\int{{\rm d}^{2\omega} k\over(2\pi)^{2\omega}}
        {k_{\mu}k_{\nu}\over [(p-k)^2-m^2][k^2-(k\cdot\eta)^2+
        x(k\cdot v)^2]^2} \,,
        \label{h5}
\end{equation}
and using~(\ref{d1}), (\ref{d2}), (\ref{d3}) and~(\ref{d4}).

\section{Gauge invariance and a Ward Identity}\label{appward}

In this appendix we prove
the Ward identity that was used in
Sect.~\ref{sec4}, i.e., $Z(v,v)=1$. We will also show
that, as a result of
gauge invariance, $q_{\mu} V^\mu_{vv'}(p,p')=0$.

Let us start with the identity $q_{\mu} V^\mu_{vv'}(p,p')=0$. This can be most
easily demonstrated by considering BRST invariance in the path integral
formalism. If $\bar c(x)$ is the anti-ghost field, we can write
 \begin{equation}
        0=\delta_{\rm BRS}\langle0|{\rm T}\psi_{v'}(y)\bar\psi_{v}(0) \bar
        c(x)|0\rangle
        \propto {\partial\over\partial x^{\mu}}\langle0|{\rm
T}\psi_{v'}(y)\bar\psi_{v}(0) A^\mu(x)
        |0\rangle \,,
        \label{e-9.3a}
 \end{equation}
where we recall that in the path integral formalism, derivatives are
understood to act outside the time ordering. Taking the Fourier
 transform of this equation, one obtains
 \begin{equation}
        0=q_{\mu}\langle \psi_{v'}(p')\bar\psi_{v}(p) A^\mu(q)
        \rangle = \ii q^\mu D_{\mu\nu}(q) \;\ii e\, V^\nu_{vv'}(p,p')
\,,
        \label{e-9.3b}
 \end{equation}
where we have used the definition~(\ref{d16}). Note that by momentum
conservation, $q=p'-p$.
Further recall that both
the free photon propagator, $D^{(0)}_{\mu\nu}(q)$, and the full
one $D_{\mu\nu}(q)$, satisfy in an arbitrary Lorentz gauge
\begin{equation}
q^\mu   D^{(0)}_{\mu\nu}(q)=q^\mu       D_{\mu\nu}(q)=-\xi{q_{\nu}\over q^2},
        \label{e-20.2a}
\end{equation}
which may be expressed as saying that the longitudinal part of the photon
propagator is not renormalised. From~(\ref{e-9.3b})
and~(\ref{e-20.2a}), and for a non-zero gauge parameter $\xi$, we
readily see that
\begin{equation}
q_{\mu} V^\mu_{vv'}(p,p')=0\,.
\label{e-9.3c}
\end{equation}
We will return to this identity, and present an alternative proof of it
using the canonical formalism, at the end of this appendix.

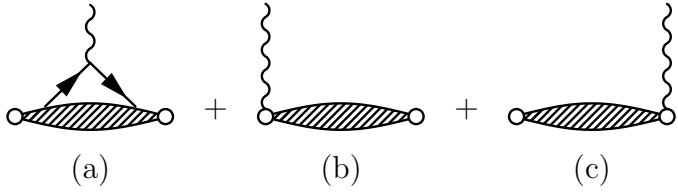
\begin{figure}[tbp]
 $$
\begin{array}{c}
\parbox[b]{20mm}{
\begin{fmfgraph*}(20,15)
\fmfstraight
\fmfbottomn{b}{2}
 \fmf{brown_muck,label=(a),label.dist=.5cm}{b1,b2}
 \fmfv{d.sh=c,d.filled=empty,d.si=2mm}{b1}
 \fmfv{d.sh=c,d.filled=empty,d.si=2mm}{b2}
 \begin{fmfsubgraph}(0.2w,0.09h)(.6w,.9h)
 \fmftop{at}
 \fmfbottomn{ab}{2}
 \fmf{fermion}{ab1,av1,ab2}
 \fmf{photon,tension=1.5}{at,av1}
 \end{fmfsubgraph}
\end{fmfgraph*}
}
\quad
+
\quad
\parbox[b]{20mm}{
\begin{fmfgraph*}(20,15)
\fmfstraight
\fmftopn{t}{2}
\fmfbottomn{b}{2}
\fmf{photon}{t1,b1}
 \fmf{brown_muck,label=(b),label.dist=.5cm}{b1,b2}
  \fmfv{d.sh=c,d.filled=empty,d.si=2mm}{b1}
 \fmfv{d.sh=c,d.filled=empty,d.si=2mm}{b2}
 \end{fmfgraph*}
}
\quad
+
\quad
\parbox[b]{20mm}{
\begin{fmfgraph*}(20,15)
\fmfstraight
\fmftopn{t}{2}
\fmfbottomn{b}{2}
\fmf{photon}{t2,b2}
 \fmf{brown_muck,label=(c),label.dist=.5cm}{b1,b2}
  \fmfv{d.sh=c,d.filled=empty,d.si=2mm}{b1}
 \fmfv{d.sh=c,d.filled=empty,d.si=2mm}{b2}
 \end{fmfgraph*}
}
\\[0.75cm]
\end{array}
$$
        \caption{Graphical representation of~(\protect\ref{e-19.2b}).
        The blobs represent irreducible contributions for the photon
        and all contributions (i.e., including
        reducible ones) for the fermions. The photon lines are amputated.}
        \label{f-19.2a}
\end{figure}

We now want to derive the Ward identity, (\ref{e-17.2c}).
Let us consider Eq.~\ref{d16} for $v'=v$. Looking at the
diagrams of Fig.~\ref{f-19.2a}, we see that we can recast the right
hand side as (we drop the ${\rm B}$ superscript
for simplicity)
  \begin{equation}
\ii e V^\nu_{vv}(p,p')=
 \ii e\langle \psi_{v}(p')\bar\psi_{v}(p)J^\nu(q)\rangle^{\rm 1PhI}
 -\ii S_{v}(p'){eV^{\nu}\over V\cdot q}+
 {eV^{\nu}\over V\cdot q}\ii S_{v}(p)
 \,,
        \label{e-19.2b}
 \end{equation}
 where 
 $J^\nu$ is the (conserved) electromagnetic current,
 $J^\nu(x)=\bar\psi(x)\gamma^\nu\psi(x)$. The superscript ``${\rm 1PhI}$''
 means one photon irreducible. A diagram is called ${\rm 1PhI}$ if it
 cannot be split into two parts by cutting a photon line.

Recalling again~(\ref{e-20.2a}),
we obtain from (\ref{e-19.2b}) and (\ref{d16})
\begin{equation}
        q_{\nu} V_{vv}^\nu(p,p')=q_{\nu}
        \langle \psi_{v}(p')\bar\psi_{v}(p)J^\nu(q)\rangle-S_{v}(p')+S_{v}(p)
        \,. \label{e-20.2b}
\end{equation}
Therefore, Eq.~\ref{e-9.3c} trivially implies
\begin{equation}
q_{\nu}
        \langle
        \psi_{v}(p')\bar\psi_{v}(p)J^\nu(q)\rangle=S_{v}(p')-S_{v}(p)\,.
        \label{e-20.2b'}
\end{equation}

We now define
\begin{equation}
{ i}e\, \mathfrak M^\mu_{vv'}(p,p')=[{ i} S_{v'}(p')]^{-1}  \,{ i}e
         \langle \psi_{v'}(p')\bar\psi_{v}(p)J^\mu(q)\rangle^{\rm 1PhI}
        [{ i} S_{v}(p)]^{-1}\,.
        \label{e-8.3a}
\end{equation}
When computing $S$-matrix elements (see \I), only diagrams with a pole
for each external leg
(here one in $p$ and one in $p'$)   are relevant, since,
in the LSZ formalism we will have to amputate such external legs.
It is thus clear that one can
construct the $S$-matrix element from (\ref{e-8.3a})  for
the scattering process since all the diagrams
with two poles are contained in the
Green's function, $\langle \psi_{v'}(p')\bar\psi_{v}(p)J^\nu(q)\rangle$.
These diagrams are contained in Fig.~\ref{f-19.2a}a; we stress that
Fig.'s~\ref{f-19.2a}b and~\ref{f-19.2a}c do not contain two poles and
cannot contribute to the $S$-matrix.

We need a renormalisation condition and, in the on-shell scheme, we impose
on the irreducible photon-dressed electron vertex
\begin{equation}
\bar u(p,s')\;\mathfrak M^{{\rm R}\,\mu}_{vv}(p,p)\;    u(p,s)=\bar u(p,s')\;
\gamma^\mu\;
        u(p,s)\,,
        \label{e-8.3b}
\end{equation}
in analogy to the well known condition
\begin{equation}
\bar u(p,s')\;\Gamma^{{\rm R}\,\mu}(p,p)\; u(p,s)
=\bar u(p,s')\;\gamma^\mu
        u(p,s)
        \,,\label{e-8.3c}
\end{equation}
for the standard QED vertex. Eq.\ \ref{e-8.3b} determines what we mean
by the physical charge, i.e., the charge that would be
experimentally measured at low momentum transfer.

It suffices, of course, to
prove that $Z(v,v)=1$ in the on-shell scheme since the UV divergent
part of the renormalisation constants is the same in any scheme, this
will also prove that $Z(v,v)=1$ in, say, the MS or
$\overline{{\rm MS}}$  schemes.

Note that from (\ref{e-8.3a}) and (\ref{e-20.2b'}) we have
\begin{equation}
 q_{\mu}\mathfrak M^\mu_{vv}(p,p')
=[S_{v}(p')]^{-1}-[S_{v}(p)]^{-1}
\,,
        \label{m11.5a}
\end{equation}
where we should point out that the dressing parameters are chosen to be
identical. In~(\ref{e-20.2b'}) we can write the expectation value as
either $\rm 1PhI$ or not, since dotting $q$ into the full photon
propagator kills everything except for the tree level propagator.
Multiplying on both sides with renormalisation constants and
sandwiching between spinors yields
\begin{equation}
 \bar u(p,s') \Zmatb{v}
 q_{\mu}\mathfrak M^\mu_{vv}(p,p') \Zmat{v} u(p,s)
= \bar u(p,s') \Zmatb{v}
\left([S_{v}(p')]^{-1}-[S_{v}(p)]^{-1}\right)
\Zmat{v} u(p,s)
\,.
        \label{m11.5b}
\end{equation}
Note now  that from the
definition of $\mathfrak M^\mu_{vv'}(p,p')$ one has
\begin{equation}
 \mathfrak M^\mu_{vv}(p,p')
= Z(v,v)\left(\Zmatb{v}\right)^{-1}\mathfrak M^{{\rm R}\,\mu}_{vv}(p,p')
\left(\Zmat{v}\right)^{-1}\,.
        \label{e-9.3e}
\end{equation}
To obtain this result we have employed~(\ref{g6}) and we further
note that the different terms on the
right hand side of~(\ref{e-19.2b}) must all renormalise in the same way since
they have completely different pole structures.
Using this and the definition of the renormalisation constants (\ref{Zmat}) for
$S_{v}(p)$,   we have
\begin{equation}
 Z(v,v)\bar u(p,s')
 q_{\mu}\mathfrak M^{{\rm R}\,\mu}_{vv}(p,p') u(p,s)
=
\bar u(p,s')\left(
[S^{R}_{v}(p')]^{-1}-[S^{R}_{v}(p)]^{-1}
\right) u(p,s)
\,.
        \label{m11.5.d}
\end{equation}
The last term on the right hand side vanishes because of the
spinor and we can similarly reexpress the first term to obtain
\begin{equation}
 Z(v,v)\bar u(p,s')
 q_{\mu}\mathfrak M^{{\rm R}\,\mu}_{vv}(p,p') u(p,s)
=
\bar u(p,s') /\kern-0.5em q u(p,s)
\,.
        \label{m11.5.e}
\end{equation}
From our renormalisation condition, the desired result, $Z(v,v)=1$,
follows.

We conclude this appendix with an alternative derivation of~(\ref{e-20.2b'}), and
hence~(\ref{e-9.3c}) for $v'=v$ using the canonical formalism.
If we notice that the left hand side of the
former is the Fourier transform of
\begin{equation}
        {\partial\over\partial x^{\nu}}\langle0|{\rm
        T}\psi_{v}(y)\bar\psi_{v}(0)J^\nu(x)|0\rangle\,,
        \label{e-20.2c}
\end{equation}
then upon taking the derivative inside the time ordered product and recalling
the conservation of electric charge, $\partial\cdot J=0$, and the
fundamental equal-time (anti-)\-commutation relations, it follows  that
\begin{equation}
{\partial\over\partial x^{\nu}}\langle0|{\rm
        T}\psi_{v}(y)\bar\psi_{v}(0)J^\nu(x)|0\rangle=
        [\delta^4(x)-\delta^4(x-y)]
        \langle0|{\rm T}\psi_{v}(y)\bar\psi_{v}(0)|0\rangle
        \,.\label{e-20.2d}
\end{equation}
In momentum space this identity can be recast as~(\ref{e-20.2b'}).
Substituting into~(\ref{e-20.2b}) we finally obtain
\begin{equation}
        q_{\nu} V_{vv}^\nu(p,p')=0\,.
        \label{e-20.2.f}
\end{equation}
The generalisation of this proof to $v'\not=v$ is straightforward.

We also note that for identical dressings ($v=v'$) the Ward
identity~(\ref{e-20.2b'}) can be understood as the usual identity in
a particular (dressing) gauge.

\end{fmffile}

\end{document}